\documentclass[twocolumn,aps,pra,floatfix]{revtex4}
\usepackage{graphicx}
\usepackage{times}
\usepackage{nicefrac}
\usepackage{amsmath}
\usepackage{amsfonts}
\usepackage{amssymb}
\usepackage{amsthm}
\usepackage{epsf}
\usepackage{bm}
\usepackage{bbm}
\usepackage{times}

\usepackage{dcolumn}

\newcolumntype{.}{D{x}{}{-1}}

\begin{document}

\newcommand{\half}{\frac12}
\newcommand{\vare}{\varepsilon}
\newcommand{\eps}{\epsilon}
\newcommand{\pr}{^{\prime}}
\newcommand{\ppr}{^{\prime\prime}}
\newcommand{\pp}{{p^{\prime}}}
\newcommand{\ppp}{{p^{\prime\prime}}}
\newcommand{\hp}{\hat{\bfp}}
\newcommand{\hr}{\hat{\bfr}}
\newcommand{\hx}{\hat{\bfx}}
\newcommand{\hpp}{\hat{\bfpp}}
\newcommand{\hq}{\hat{\bfq}}
\newcommand{\rqq}{{\rm q}}
\newcommand{\bfk}{{\bm{k}}}
\newcommand{\bfp}{{\bm{p}}}
\newcommand{\bfq}{{\bm{q}}}
\newcommand{\bfr}{{\bm{r}}}
\newcommand{\bfx}{{\bm{x}}}
\newcommand{\bfy}{{\bm{y}}}
\newcommand{\bfz}{{\bm{z}}}
\newcommand{\bfpp}{{\bm{\pp}}}
\newcommand{\bfppp}{{\bm{\ppp}}}
\newcommand{\balpha}{\bm{\alpha}}
\newcommand{\bvare}{\bm{\vare}}
\newcommand{\bgamma}{\bm{\gamma}}
\newcommand{\bGamma}{\bm{\Gamma}}
\newcommand{\bLambda}{\bm{\Lambda}}
\newcommand{\bmu}{\bm{\mu}}
\newcommand{\bnabla}{\bm{\nabla}}
\newcommand{\bvarrho}{\bm{\varrho}}
\newcommand{\bsigma}{\bm{\sigma}}
\newcommand{\bTheta}{\bm{\Theta}}
\newcommand{\bphi}{\bm{\phi}}
\newcommand{\bomega}{\bm{\omega}}
\newcommand{\intzo}{\int_0^1}
\newcommand{\intinf}{\int^{\infty}_{-\infty}}
\newcommand{\lbr}{\langle}
\newcommand{\rbr}{\rangle}
\newcommand{\ThreeJ}[6]{
        \left(
        \begin{array}{ccc}
        #1  & #2  & #3 \\
        #4  & #5  & #6 \\
        \end{array}
        \right)
        }
\newcommand{\SixJ}[6]{
        \left\{
        \begin{array}{ccc}
        #1  & #2  & #3 \\
        #4  & #5  & #6 \\
        \end{array}
        \right\}
        }
\newcommand{\NineJ}[9]{
        \left\{
        \begin{array}{ccc}
        #1  & #2  & #3 \\
        #4  & #5  & #6 \\
        #7  & #8  & #9 \\
        \end{array}
        \right\}
        }
\newcommand{\Vector}[2]{
        \left(
        \begin{array}{c}
        #1     \\
        #2     \\
        \end{array}
        \right)
        }

\newcommand{\Dmatrix}[4]{
        \left(
        \begin{array}{cc}
        #1  & #2   \\
        #3  & #4   \\
        \end{array}
        \right)
        }
\newcommand{\Dcase}[4]{
        \left\{
        \begin{array}{cl}
        #1  & #2   \\
        #3  & #4   \\
        \end{array}
        \right.
        }
\newcommand{\cross}[1]{#1\!\!\!/}

\newcommand{\Za}{{Z \alpha}}
\newcommand{\im}{{ i}}

\title{Two-loop self-energy in the Lamb shift of the ground and excited states of hydrogen-like ions}

\author{V.~A. Yerokhin}
\affiliation{Center for Advanced Studies, Peter the Great St.~Petersburg Polytechnic University,
195251 St.~Petersburg, Russia}

\begin{abstract}

The two-loop self-energy correction to the Lamb shift of hydrogen-like ions is calculated for the
$1s$, $2s$, and $2p_{1/2}$ states and nuclear charge numbers $Z = 30$-$100$. The calculation is
performed to all orders in the nuclear binding strength parameter $\Za$. As compared to previous
calculations of this correction, numerical accuracy is improved by an order of magnitude and the
region of the nuclear charges is extended. An analysis of the $Z$-dependence of the obtained
results demonstrates their consistency with the known $\Za$-expansion coefficients.
\end{abstract}

\maketitle

\section{Introduction}

Theoretical and experimental investigations of the Lamb shift in atomic systems provide stringent
tests of the bound-state quantum electrodynamics (QED) through the first two orders in the
fine-structure constant $\alpha$ and to all orders in the nuclear binding strength parameter $\Za$
(where $Z$ is the nuclear charge number) \cite{mohr:98,shabaev:02:rep}. The main factors presently
limiting our theoretical understanding of the Lamb shift in hydrogen-like atoms are
\cite{yerokhin:15:Hlike} the binding two-loop QED effects and, in particular, the two-loop
self-energy. Accurate treatment of the two-loop effects is crucial for an adequate comparison of
theory and experiment along the whole range of nuclear charge numbers, from hydrogen
\cite{fischer:04} till lithium-like uranium \cite{brandau:04,beiersdorfer:05}.

The two-loop QED effects were extensively investigated during the last decades, both within the
method based on the $\Za$ expansion
\cite{pachucki:01:pra,pachucki:03:prl,czarnecki:05:prl,jentschura:05:sese} and also within the
all-order (in $\Za$) approach
\cite{yerokhin:01:prl,yerokhin:03:prl,yerokhin:06:prl,yerokhin:08:twoloop,yerokhin:09:sese}.
Despite significant progress achieved in these studies, at least two important issues remains to be
solved. The first is that the extrapolation of the all-order results
\cite{yerokhin:05:sese,yerokhin:09:sese} for the $1s$ two-loop self-energy towards $Z\to 0$ is only
barely consistent with the $\Za$ expansion results \cite{pachucki:03:prl}. The associated
uncertainty is presently the largest theoretical error for the hydrogen Lamb shift
\cite{mohr:16:codata}. In the foreseeable future (once the proton charge radius puzzle
\cite{pohl:05} is solved), this error will define the uncertainty of the Rydberg constant, which is
determined  \cite{mohr:16:codata} from the hydrogen spectroscopy. The second issue is that the
calculation of Ref.~\cite{yerokhin:06:prl} of the two-loop self-energy for the $n = 2$ states was
performed only for several ions with $Z \ge 60$. An extension of these calculations towards lower
values of $Z$ is needed.

In the present work we report a calculation of the two-loop self-energy correction for the $1s$,
$2s$, and $2p_{1/2}$ states of hydrogen-like ions. As compared to previous calculations, we enhance
numerical accuracy by an order of magnitude and extend calculations for the $n = 2$ states to lower
values of $Z$, till $Z = 30$. We also present a detailed description of the method of calculation,
as developed during two decades of our work on this problem. In our 2003 paper
\cite{yerokhin:03:epjd} we already reported a detailed analysis of the two-loop self-energy and
described the basic scheme of the calculation to all orders in $\Za$. The problem in hand is rather
complex and its complete description would be unnecessary long; for this reason, in the present
work we will concentrate mainly on new features of the calculational method and only sketch the
parts that can be found in Ref.~\cite{yerokhin:03:epjd}.

In the present work we will use the Feynman gauge for the photon propagator since this will make
formulas more compact. The actual calculation will be performed for the point distribution of the
nuclear charge, although the general analysis will be valid for other nuclear-charge distributions
as well. Notations and definitions used throughout the paper are collected in Appendix~\ref{app:1}.
Appendices~\ref{app:2} and \ref{app:se} contain basic notations and formulas for the operator of
the electron-electron interaction and the one-loop self-energy, which are essential for this work.
We use the relativistic units ($\hbar=c=1$) and the Heaviside charge units ($ \alpha = e^2/4\pi$,
$e<0$).

\section{Basic approach}

\begin{figure}
\centerline{
\resizebox{\columnwidth}{!}{%
  \includegraphics{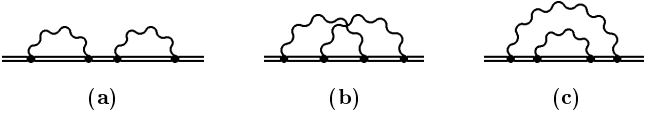}
} }
\vskip0.5cm
\caption{Two-loop self-energy diagrams. The double line denotes the electron propagating in the
binding Coulomb field of the nucleus. The wave lines denote virtual photons.
Individual graphs are referred to as
the loop-after-loop (LAL) diagram (a), the overlapping diagram (b), and the
nested diagram (c).  \label{fig:sese}}
\end{figure}

The two-loop self-energy correction is represented by Feynman diagrams in Fig.~\ref{fig:sese}. The
corresponding formal expression can be easily derived, e.g., by the two-time Green function method
\cite{shabaev:02:rep},
\begin{align} \label{eq1}
\Delta E_{\rm SESE} = \Delta E_{\rm LAL} + \Delta E_{N}  + \Delta E_{O} + \Delta E_{\rm red}\,.
\end{align}
The first term in the right-hand-side of Eq.~(\ref{eq1}) is the loop-after-loop (LAL) correction
induced by the irreducible ($n\neq a$) part of the diagram in Fig.~\ref{fig:sese}(a),
\begin{align}
\Delta E_{\rm LAL} = \sum_{n\neq a} \frac{\lbr a|\gamma^0\widetilde{\Sigma}(\vare_a)|n\rbr \lbr n|\gamma^0\widetilde{\Sigma}(\vare_a)|a\rbr }{\vare_a-\vare_n}\,,
\end{align}
where $a$ denotes the reference state, the summation over $n$ is performed over the spectrum of the
Dirac-Coulomb Hamiltonian, $\vare_a$ and $\vare_n$ are the Dirac-Coulomb energy eigenvalues of the
states $a$ and $n$, respectively, and $\widetilde{\Sigma}(\vare) = \Sigma(\vare)-\delta m$ is the
one-loop self-energy operator described in Appendix~\ref{app:se}. The LAL correction can be
evaluated by using generalizations of approaches developed for the one-loop self-energy. Such
calculations were performed by several groups
\cite{mitrushenkov:95,mallampalli:98:prl,yerokhin:00:lalpra}. Since this part of the calculation is
relatively straightforward  and well established, it will not be described here.

The last three terms in the right-hand-side of Eq.~(\ref{eq1}) comprise the nontrivial part of the
two-loop self-energy, which is the main subject of the present work. The contribution of the
overlapping ($O$) diagram in Fig.~\ref{fig:sese}(b) reads
\begin{align}  \label{eq3O}
\Delta E_{O} &\ =  (2i\alpha)^2
   \int_{C_F} d\omega_1\,d\omega_2 \int d\bfx_1\ldots d\bfx_4\,
    D(\omega_1,x_{13})\,
\nonumber \\ &  \times
    D(\omega_2,x_{24})\,
   {\psi}^{\dag}_a(\bfx_1)\, \alpha_{\mu}\, G(\vare_a-\omega_1,\bfx_1,\bfx_2)\,
           \alpha_{\nu}\,
\nonumber \\ &  \times
  G(\vare_a-\omega_1-\omega_2,\bfx_2,\bfx_3)\,\alpha^{\mu}\, G(\vare_a-\omega_2,\bfx_3,\bfx_4)\,
\nonumber \\ &  \times
         \alpha^{\nu}
     \psi_a(\bfx_4)\,,
\end{align}
where $x_{ij} = |\bfx_i-\bfx_j|$, $\psi_a(\bfx)$ is the reference-state wave function,
$D(\omega,\bm{x})$ is the scalar part of the photon propagator, $G(\vare,\bfx_i,\bfx_j)$ is the
Dirac-Coulomb Green function (see Appedix~\ref{app:1} for notations and definitions), and $C_F$ is
the standard Feynman integration contour. The contribution of the nested ($N$) diagram in
Fig.~\ref{fig:sese}(c) is
\begin{align}  \label{eq4}
\Delta E_{N} &\ = (2i\alpha)^2 \int_{C_F} d\omega_1\,d\omega_2\,
   \int d\bfx_1 \ldots d\bfx_4\,
    D(\omega_1,x_{14})\,
\nonumber \\ &  \times
    D(\omega_2,x_{23})\,
       {\psi}^{\dag}_a(\bfx_1)\, \alpha_{\mu}
          G(\vare_a-\omega_1,\bfx_1,\bfx_2) \, \alpha_{\nu}\,
\nonumber \\ &  \times
           G(\vare_a-\omega_1-\omega_2,\bfx_2,\bfx_3)\, \alpha^{\nu}\,
          G(\vare_a-\omega_1,\bfx_3,\bfx_4)\,
\nonumber \\ & \times
          \alpha^{\mu}\, \psi_a(\bfx_4)\,.
\end{align}
The fourth term in Eq.~(\ref{eq1}) is the contribution of the reducible ($n = a$) part of the
diagram in Fig.~\ref{fig:sese}(a),
\begin{align}  \label{eq5}
\Delta E_{\rm red} = \Delta E_{\rm SE}\, \lbr a| \left. \gamma^0
    \frac{\partial}{\partial \vare} \Sigma (\vare) \right|_{\vare=\vare_a}
            |a\rbr\,,
\end{align}
where $\Delta E_{\rm SE} \equiv \lbr a|\gamma^0\widetilde{\Sigma}(\vare_a)|a\rbr $ is the one-loop
self-energy correction described in Appendix~\ref{app:se}.

\begin{figure}
\centerline{\resizebox{0.5\textwidth}{!}{\includegraphics{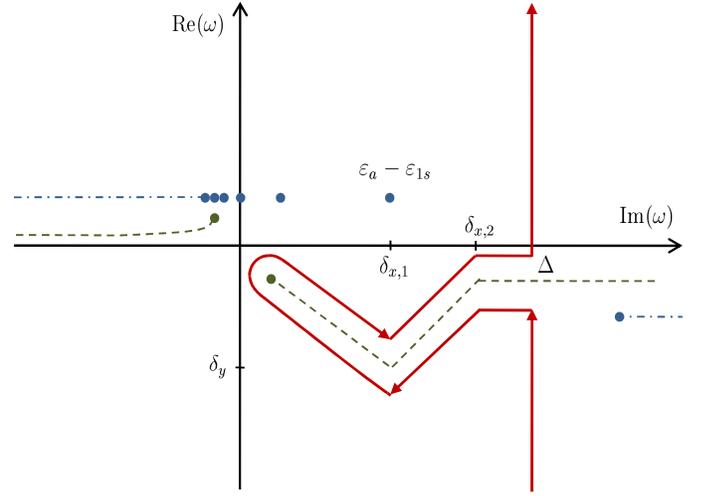}}}
 \caption{The integration contour $C_{LH}$ in the complex $\omega$ plane. The branch cuts of the photon
 propagator are shown with the dashed line (green). The poles and the branch cuts of the electron propagator
 are shown by dots and the dashed-dot line (blue). The poles and branch cuts are shown for the case of
 the one-loop self-energy of an excited state. \label{fig:CLH}}
\end{figure}

Equations (\ref{eq3O})-(\ref{eq5}) are only formal expressions and need to be renormalized and
reformulated, in order to be made suitable for a numerical evaluation. In particular, all
ultraviolet (UV) divergences should be regularized in a covariant way and explicitly cancelled. The
main problem of the renormalization originates from the fact that the UV divergences are usually
identified and cancelled in momentum space, whereas the Dirac-Coulomb Green function is known in
the coordinate space and, moreover, in the form of the partial-wave expansion only. There are
studies that perform the renormalization of the bound-state QED corrections in coordinate space
\cite{indelicato:92:se,indelicato:98,indelicato:01,indelicato:14}, but they are restricted to the
one-loop level so far.

In order to covariantly regularize and identify UV divergences present in
Eqs.~(\ref{eq3O})-(\ref{eq5}), we use the approach based on the expansion of the Dirac-Coulomb
Green function  $G$ in terms of interactions with the binding Coulomb field $V$,
\begin{align} \label{eq6}
G  &\ = G^{(0)}  + G^{(1)}  + G^{(2+)}
 \nonumber \\ &
\equiv G^{(0)}
+ G^{(0)} \,V\,G^{(0)}
+ G^{(0)} \,V\,G^{(0)} \,V\,G
\,,
\end{align}
where $G^{(0)}$ is the free Dirac Green function.

In the case of the one-loop self-energy, the renormalization approach is quite straightforward
\cite{snyderman:91}. The UV divergences are induced only by the first two terms of expansion
(\ref{eq6}). These terms contain only the free Dirac Green function $G^{(0)}$ and are evaluated in
momentum space. The remainder does not contain any UV divergences and is evaluated in coordinate
space.

For the two-loop self-energy, the renormalization procedure becomes more complicated, mainly
because of appearance of mixed terms containing both the UV-divergent subgraphs and the
Dirac-Coulomb Green function. The two-loop renormalization procedure based on expansion (\ref{eq6})
was first sketched by Mallampalli and Sapirstein \cite{mallampalli:98:pra} and fully realized in
our investigations \cite{yerokhin:01:sese,yerokhin:03:epjd}. Following these studies, we split each
of the nested, overlapping, and reducible contribution in Eq.~(\ref{eq1}) into three parts, which
are termed as the $M$, $P$, and $F$ terms, correspondingly,
\begin{align}  \label{eq6b}
\Delta E_{O} =&\ \Delta E_{O,M} + \Delta E_{O,P} + \Delta E_{O,F}\,,\\
\Delta E_{N} =&\ \Delta E_{N,M} + \Delta E_{N,P} + \Delta E_{N,F}\,,\\
\Delta E_{{\rm red}} =&\ \Delta E_{{\rm red},M} + \Delta E_{{\rm red},P} + \Delta E_{{\rm red},F}\,.
\end{align}
The $M$ terms are free from any UV divergences; they are evaluated in coordinate space by using the
partial-wave expansions of the Dirac-Coulomb Green functions. The $F$ terms contain only the free
Dirac Green functions; they are evaluated in momentum space, without any partial-wave expansions.
The $P$ terms contain both the Dirac-Coulomb Green functions {\em and} one-loop UV-divergent
subgraphs. They are evaluated in the mixed representation: the UV-divergent subgraphs are treated
in momentum space, whereas the Dirac-Coulomb Green functions are represented as a Fourier transform
over one of the radial variables.

It is convenient to group the $M$, $P$, and $F$ terms together, thus representing the two-loop
self-energy correction as
\begin{align}
\Delta E_{\rm SESE} = \Delta E_{\rm LAL} + \Delta E_{M}  + \Delta E_{P} + \Delta E_{F}\,,
\end{align}
where
\begin{align}  \label{eq6c}
\Delta E_{M} =&\ \Delta E_{O,M} + \Delta E_{N,M} + \Delta E_{{\rm red},M}\,,\\
\Delta E_{P} =&\ \Delta E_{O,P} + \Delta E_{N,P} + \Delta E_{{\rm red},P}\,,\\
\Delta E_{F} =&\ \Delta E_{O,F} + \Delta E_{N,F} + \Delta E_{{\rm red},F}\,.
\label{eq6d}
\end{align}
The exact definitions of the $M$, $P$, and $F$ terms are given in the next three sections.

In our previous investigation \cite{yerokhin:03:epjd} we presented a detailed analysis of the
divergences arising in individual contributions in Eqs.~(\ref{eq6c})-(\ref{eq6d}) and demonstrated
their cancellation. In the present work we will rely on that analysis and assume all UV divergences
in individual contributions to be renormalized, i.e., divergent terms $\sim\!\!1/\eps$ and
$\sim\!\!1/\eps^2$ in $D = 4-2\eps$ dimensions will be dropped out. The reference-state infrared
(IR) divergences will be removed from each individual contribution by IR subtractions. The net
effect of all IR subtractions is zero, which follows from the analysis of
Ref.~\cite{yerokhin:03:epjd}. So, in our present formulation each of the individual contributions
in the right-hand-side of Eqs.~(\ref{eq6c})-(\ref{eq6d}) is finite. This fact will greatly simplify
the following description.

%
%
%
\section{${\bf M}$ term}
\label{sec:mterm}

The nested, overlapping and reducible $M$ terms are obtained from Eqs.~(\ref{eq3O})-(\ref{eq5}) by
applying subtractions that remove all UV and reference-state IR divergences. Each of these terms in
turn will be discussed in the next three subsections. In order to keep formulas compact, we will
have to repeatedly switch between the Green-function form and the spectral representation of the
electron propagators.

\begin{widetext}

\subsection{Nested $\bm{M}$ term}

The nested $M$ term is given by
\begin{align} \label{eqMN1}
 \Delta E_{N,M} &\ = \left( \frac{i}{2\pi}\right)^2
   \int_{C_{F}} d\omega_1\, d\omega_2\,
         \Biggl[ \sum_{n_1n_2n_3}
    \frac{\lbr an_3|I(\omega_1)|n_1a \rbr \lbr n_1n_2|I(\omega_2)|n_2n_3\rbr}
      {(\vare_a-\omega_1-\vare_{n_1})(\vare_a-\omega_1-\omega_2-\vare_{n_2}) (\vare_a-\omega_1-\vare_{n_3})
         }
    \nonumber \\ &
    -
    \frac1{\omega_1^2} \sum_{a'a''}\lbr aa''|I(\omega_1)|a'a\rbr
    \sum_{n_2}\frac{\lbr a'n_2|I(\omega_2)|n_2a''\rbr}
      {\vare_a-\omega_2-\vare_{n_2}    }
- \mbox{UV subtraction}
      \Biggr]    \,,
\end{align}
where $I(\omega)$ is the operator of the electron-electron interaction defined in
Appendix~\ref{app:2}, the summation over $n_i$ is performed over the complete Dirac spectrum, and
intermediate states $a'$ and $a''$ differ from the reference state $a$ only by the momentum
projection. The first term in the brackets in the above formula is the unsubtracted nested
contribution given by Eq.~(\ref{eq4}). The second term in brackets is the IR subtraction that
cancels the reference-state IR divergence present in the first term. The third term in the brackets
is the UV subtraction that is schematically represented by the following substitution (to be
applied both to the first and second terms in the brackets)
\begin{align}
 G_2(\vare) \equiv \sum_{n_2} \frac{|n_2\rbr\lbr n_2|}{\vare-\vare_{n_2}}\  &\to  G_2^{(2+)}(\vare)\,,
\end{align}
where $G^{(2+)}(\vare)$ is the Dirac Green function containing two and more interactions with the
binding Coulomb field (see Appendix~\ref{app:1} for the exact definition).

The IR divergence in the unsubtracted nested contribution was discussed in detail in
Ref.~\cite{yerokhin:03:epjd}. It was shown that it appears in the limit $\omega_1\to 0$ when both
the $n_1$ and $n_3$ intermediate states are degenerate in energy with the reference state $a$,
$\vare_{n_1} = \vare_{n_2} = \vare_{a}$. We now add that the divergences arise only when the $n_1$
and $n_3$ intermediate states have the same parity as the reference state $a$. In particular, for
the $a = 2s$ reference state, the contribution of the $n_1 = n_3 = 2p_{1/2}$ intermediate states is
IR finite, since the radial integrals in the numerator vanish in the limit $\omega_1\to 0$ due to
orthogonality of the wave functions. Therefore, the IR divergences originate from the intermediate
states $n_1 = a'$ and $n_2 = a''$ that may differ from the reference state $a$ only by the momentum
projection ($\mu_{a'}$ and $\mu_{a''}$, respectively). Hence, the the sum over $a'$ and $a''$ in
Eq.~(\ref{eqMN1}) is actually the sum over all possible values of $\mu_{a'}$ and $\mu_{a''}$.

In order to bring Eq.~(\ref{eqMN1}) to a form suitable for a numerical evaluation, we need to sum
over the magnetic substates, perform integrations over the angular variables, and deform the
contour of the $\omega_1$ and $\omega_2$ integrations. In the present work we use the same
integration contour $C_{LH}$ as for the one-loop self-energy, described in Appendix~\ref{app:se}
and shown in Fig.~\ref{fig:CLH}. This contour is similar to the one introduced by P.~Mohr in
Ref.~\cite{mohr:74:a} but differs in details. The fact that we can deform the original contour
$C_F$ to $C_{LH}$ in the two-loop self-energy follows from the analysis presented in Appendix B of
Ref.~\cite{yerokhin:03:epjd}. The resulting expression for the nested $M$ term is
\begin{align}  \label{eqMN2}
\Delta E_{N,M} &\ = \left( \frac{i\alpha}{2\pi}\right)^2
   \int_{C_{LH}} d\omega_1\, d\omega_2\, \Biggl[ \sum_{{n_1n_2n_3}\atop{J_1J_2}}
     \frac{(-1)^{J_1+J_2} X_N\,
    R_{J_1}(\omega_1,an_3n_1a)\,
      R_{J_2}(\omega_2,n_1n_2n_2n_3)}
      {(\vare_a-\omega_1-\vare_{n_1})(\vare_a-\omega_1-\omega_2-\vare_{n_2})(\vare_a-\omega_1-\vare_{n_3})}
\nonumber \\ &
        -
    \sum_{J_1} \frac{(-1)^{J_1}R_{J_1}(\omega_1,aaaa)}{\omega_1^2\,(2j_a+1)^2}\,
            \sum_{n_2J_2}\frac{ (-1)^{j_2-j_a+J_2}\, R_{J_2}(\omega_2,an_2n_2a)} {\vare_a-\omega_2-\vare_{n_2}    }
    - \mbox{UV subtraction}
      \Biggr]    \,,
\end{align}
where $R_J(\omega,abcd)$ is the relativistic generalization of the Slater radial integral (see
Appendix~\ref{app:2}), $X_N$ is the angular coefficient given by
\begin{align}
X_N = \frac{(-1)^{j_2-j_a} \delta_{\kappa_1 \kappa_3}}{(2j_a+1)(2j_1+1)} \,,
\end{align}
$j$ denotes the total angular momentum, and $\kappa$ is the relativistic angular quantum number of
the corresponding electron state.

The expression (\ref{eqMN2}) is finite and can be evaluated numerically as it stands. For the
convenience of the numerical computation, however, we divide it in three parts and evaluate each of
them separately. First, we single out the contribution with $n_1 = n_2 = n_3 = a$ from the first
term in brackets of Eq.~(\ref{eqMN2}), together with the corresponding part ($n_2 = a$) of the IR
subtraction term. The sum of them is finite but nearly divergent. It can be transformed to a more
regular form by performing the integrations over $\omega_1$ and $\omega_2$, as illustrated in
Sec.~1.3 of Ref.~\cite{yerokhin:03:epjd}. The result is
\begin{align} \label{infr2a}
\Delta E_{N,M,a} &\ =
 \frac1{(2j_a+1)^2}
  \frac{\alpha^2}{\pi^2}
\int_0^{\infty} dk_1\, dk_2\,
        \frac{ 1}{k_1k_2(k_1+k_2)}\,
 {\rm Im}\Bigl[\sum_{J_1} (-1)^{J_1} R_{J_1}(k_1,aaaa)\Bigr]\,
 {\rm Im}\Bigl[\sum_{J_2} (-1)^{J_2} R_{J_2}(k_2,aaaa)\Bigr] \,.
\end{align}

Second, we separate the contribution with $n_1 = n_3 = a$ (but $n_2 \ne a$) from the first term in
brackets of Eq.~(\ref{eqMN2}), together with the remaining part ($n_2 \ne a$) of the IR subtraction
term, and the corresponding part of the UV subtractions. Combining them together, we obtain
\begin{align}  \label{eqMN3}
\Delta E_{N,M,i}  = \left( \frac{i\alpha}{2\pi}\right)^2
   \int_{C_{LH}} d\omega_1\, d\omega_2\, &\
 \Biggl[
    \sum_{J_1} \frac{(-1)^{J_1}R_{J_1}(\omega_1,aaaa)}{\omega_1^2\,(2j_a+1)^2}\,
            \sum_{{n_2\ne a}\atop{J_2}}
            (-1)^{j_2-j_a+J_2}\, R_{J_2}(\omega_2,an_2n_2a)
 \nonumber \\ & \times \left( \frac1{\vare_a-\omega_1-\omega_2-\vare_{n_2}}
            -\frac1{\vare_a-\omega_2-\vare_{n_2}}\right)
    - \mbox{UV subtraction}
      \Biggr]    \,.
\end{align}
This part yields the dominant numerical contribution in the low-$Z$ region but is relatively simple
to evaluate and it contains only one partial-wave summation. Finally, the remainder of
Eq.~(\ref{eqMN2}) is denoted as $\Delta E_{N,M,r}$ and is evaluated separately. This part contains
two partial-wave summations and its computation is rather complicated and time consuming. The
advantage of evaluating separately $\Delta E_{N,M,r}$ is that it does not suffer (too much) from
numerical cancellations occurring in the low-$Z$ region.

\subsection{Overlapping $\bm{M}$ term}

The overlapping $M$ term is can be written as
\begin{align} \label{eqMO1}
&\ \Delta E_{O,M}  = \left( \frac{i}{2\pi}\right)^2
   \int_{C_{F}} d\omega_1\, d\omega_2\,
  \Biggl[ \sum_{n_1n_2n_3}
    \frac{\lbr an_2|I(\omega_1)|n_1n_3 \rbr \lbr n_1n_3|I(\omega_2)|n_2a\rbr}
      {(\vare_a-\omega_1-\vare_{n_1})(\vare_a-\omega_1-\omega_2-\vare_{n_2})(\vare_a-\omega_2-\vare_{n_3})}
- \, \mbox{\rm UV subtractions} \Biggr]\,,
\end{align}
where the first term in the brackets is the unsubtracted overlapping contribution as given by
Eq.~(\ref{eq3O}) and the UV subtractions are schematically represented by
\begin{align}
 G_1G_2G_3\  &\to  G_1G_2G_3 - G_1G_2^{(0)}G_3^{(0)}
 -G_1^{(0)}G_2^{(0)}G_3
    + G_1^{(0)}G_2^{(0)}G_3^{(0)} -G_1^{(0)} G_2^{(1)} G_3^{(0)} \,.
\end{align}
Here, the index of $G$ corresponds to the index of $n$ in Eq.~(\ref{eqMO1}), i.e., $G_i(\vare)
\equiv \sum_{n_i}|n_i\rbr \lbr n_i|/(\vare-\vare_{n_i})$.

Summing over the magnetic substates, performing integrations over the angular variables, and
deforming the integration contour of the $\omega_1$ and $\omega_2$ integrations, we obtain
\begin{align}\label{eqMO3}
\Delta
E_{O,M} &\ = \left( \frac{i\alpha}{2\pi}\right)^2
   \int_{C_{LH}} d\omega_1\, d\omega_2\,
 \Biggl[  \sum_{{n_1n_2n_3}\atop{J_1J_2}}
    \frac{X_O^{J_1J_2} R_{J_1}(\omega_1,an_2n_1n_3)\, R_{J_2}(\omega_2,n_1n_3n_2a)}
      {(\vare_a-\omega_1-\vare_{n_1})(\vare_a-\omega_1-\omega_2-\vare_{n_2})(\vare_a-\omega_2-\vare_{n_3})}
- \, \mbox{\rm UV subtractions} \Biggr]\,,
\end{align}
\end{widetext}
where
\begin{align}
X_O^{J_1J_2} = \frac{(-1)^{j_1+j_2+j_3-j_a}}{2j_a+1}
\SixJ{j_2}{J_2}{j_1}{j_a}{J_1}{j_3}\,.
\end{align}

Equation (\ref{eqMO3}) is finite and can be evaluated numerically as it stands. For the convenience
of the numerical computation, however, we separate it in several parts and evaluate them
separately.

First, we single out the contribution with $n_1 = n_2 = n_3 = a$ and transform it to a more regular
form by evaluating the integrations over $\omega_1$ and $\omega_2$ (see Sec.~1.3 of
Ref.~\cite{yerokhin:03:epjd}), with the result
\begin{align}
&\ \Delta E_{O,M,a}  = -\frac{\alpha^2}{\pi^2} \int_0^{\infty} dk_1\,
    dk_2\,  \frac1{k_1k_2(k_1+k_2)}
\nonumber \\ & \times
    \sum_{J_1J_2} X_O^{J_1J_2}\,
 {\rm Im}\Bigl[ R_{J_1}(k_1,aaaa)\Bigr]\,
{\rm Im}\Bigl[R_{J_2}(k_2,aaaa)\Bigr] \,.
\end{align}

Second, we separate out the part of $\Delta E_{O,M}$ that is relatively simple, contains only one
partial-wave summation, but yields the dominant numerical contribution in the low-$Z$ region,
$\Delta E_{O,M,i}$. We define it as a part of the right-hand-side of Eq.~(\ref{eqMO3}) with the
following restrictions: $\kappa_1 = \kappa_a$, $\kappa_2 = \kappa_3$, $R_{J_1} \to R^{C}_0$ (where
$R^C_J $ is the Coulomb part of $R_{J}$), and the symmetrical contribution with $\kappa_1 =
\kappa_2$, $\kappa_3 = \kappa_a$, and $R_{J_2} \to R^{C}_0$. The remaining part, $\Delta
E_{O,M,r}$, contains two unbound partial wave summations and is evaluated separately. Its
computations is the most complicated and time-consuming part of the calculation of the $M$ term.

\subsection{Reducible $\bm{M}$ term}

The reducible $M$ term is given by
\begin{align}  \label{eqMred}
&\ \Delta E_{{\rm red},M} = \Delta E_{\rm SE}\,\left(-\frac{i\alpha}{2\pi}\right)\int_{C_F}d\omega\,
 \nonumber \\ & \times
\Biggl[ \sum_n \frac{\lbr an|I(\omega)|na\rbr}{(\vare_a-\omega-\vare_n)^2}
-\sum_{{a'}} \frac{\lbr aa'|I(\omega)|a'a\rbr}{(-\omega)^2}
 \nonumber \\ &
- \sum_{\alpha } \frac{\lbr a\alpha |I(\omega)|\alpha a\rbr}{(\vare_a-\omega-\vare_{\alpha })^2} \Biggr]\,,
\end{align}
where $\Delta E_{\rm SE}$ is the one-loop self-energy correction to the energy (see
Appendix~\ref{app:se}) and the summation over $\alpha$ is performed over the spectrum of the {\em
free} Dirac Hamiltonian. The first term in the brackets of the above formula corresponds to the
unsubtracted reducible term of Eq.~(\ref{eq5}), the second term is the IR subtraction, and the
third term is the free-electron ($Z = 0$) UV subtraction, $G^{(0)}(\vare) \equiv \sum_{\alpha
}|\alpha \rbr\lbr \alpha |/(\vare-\vare_{\alpha })$. Expression (\ref{eqMred}) can be evaluated
after deforming the integration contour $C_F \to C_{LH}$. Its computation is relatively
straightforward and was performed by adapting the general scheme developed for the one-loop
self-energy matrix element \cite{yerokhin:05:se}.

\subsection{Numerical evaluation of $\bm{M}$ terms}

The general scheme of our numerical computation of the $M$ terms was described in detail in
Ref.~\cite{yerokhin:03:epjd} and does not need to be repeated here. We therefore will concentrate
on new features of our computational method. One of the important differences introduced in the
present work was the choice of the integration contour for the $\omega_1$ and $\omega_2$
integrations. We now use the contour $C_{LH}$ as described in Appendix~\ref{app:se}, which has
several advantages as compared to the standard Wick rotation ($\omega \to i\omega$) employed  in
Ref.~\cite{yerokhin:03:epjd}. First, by bending the low-energy part of the contour into the complex
plane, we avoid the appearance of numerous pole terms, which significantly simplifies the analysis
in the case of excited states. Second, this choice of the contour softens the infrared
(small-$\omega$) behaviour of the integrand, due to appearance of $\sin(\omega r_{12})$ from the
photon propagators, instead of $\exp(-\omega r_{12})$ for the Wick rotation. Because of this, the
integrand has a much more regular behaviour at small $\omega_1$ and (or) $\omega_2$, which
significantly simplifies numerical integrations for low values of $Z$.

Calculations of the $\Delta E_{N,M,i}$ and $\Delta E_{O,M,i}$ parts involve only a single
partial-wave expansion over the relativistic angular momentum parameter $\kappa$. It was extended
up to $|\kappa_{\rm max}| = 20$--$22$ and the tail of the expansion was estimated by fitting the
expansion terms to the polynomials in $1/|\kappa|$.

Calculations of $\Delta E_{N,M,r}$ and $\Delta E_{O,M,r}$ involve a double partial-wave expansion
over $\kappa$'s of two (out of the three) electron propagators; we chose them to be $\kappa_1$ and
$\kappa_3$. An important issue is the extrapolation of $|\kappa_1|\to \infty$ and $|\kappa_3|\to
\infty$. Following Ref.~\cite{yerokhin:03:epjd}, instead of calculating the matrix of the
partial-wave contributions $X_{|\kappa_1|,|\kappa_3|}$ and then extrapolating it, we prefer to work
with the matrix $Y_{l_1 l_2}$, where $l_1 = ||\kappa_1|-|\kappa_3|| = 0,1,\ldots$ is the
consecutive number of the subdiagonal in the matrix $X_{|\kappa_1|,|\kappa_3|}$ and $l_2 =
(|\kappa_1|+|\kappa_3|-l_1)/2 = 1,2,\ldots$ is the consecutive number of the element in the $l_1$th
subdiagonal of $X_{|\kappa_1|,|\kappa_3|}$. In our computations, we compute the elements of the
matrix $Y_{l_1l_2}$ up to $(l_{1,\rm max},l_{2,\rm max}) = (14,10)$ for $Z = 30,40,$ and $50$, and
$(12,10)$ for higher values of $Z$. We note that for the overlapping diagram
$X_{|\kappa_1|,|\kappa_3|}$ = $X_{|\kappa_3|,|\kappa_1|}$, which reduces the number of matrix
elements to be computed. The extrapolation was performed in two steps. First, we extrapolate
$l_2\to \infty$ by fitting the expansion terms to the polynomials in $1/l_2$. Second, we
extrapolate $l_1\to \infty$ in a similar way.

Numerical results for the $M$ terms for the $1s$, $2s$, and $2p_{1/2}$ states of hydrogen-like ions
with $Z = 30$-$100$ are presented in Table~\ref{tab:mterm}. The results for the $1s$ state are in
agreement with and more accurate than our previous values reported in Ref.~\cite{yerokhin:05:sese}.
The results for the $2s$ and $2p_{1/2}$ states extend our previous calculations reported in
Ref.~\cite{yerokhin:06:prl} and improve their accuracy by an order of magnitude.

\begin{table*}
\caption{Numerical results for the $M$ contribution $\Delta E_{M}$, in units of $F(\Za)$
 defined in Eq.~(\ref{FZa}). \label{tab:mterm}}
\begin{ruledtabular}
\begin{tabular}{ldddddd}
 \multicolumn{1}{c}{$Z$}   & \multicolumn{1}{c}{$\Delta E_{{\rm red}, M}$}
    & \multicolumn{1}{c}{$\Delta E_{N,M,i+a}$}
    & \multicolumn{1}{c}{$\Delta E_{N,M,r}$}
    & \multicolumn{1}{c}{$\Delta E_{O,M,i+a}$}  & \multicolumn{1}{c}{$\Delta E_{M,O,r}$}  & \multicolumn{1}{c}{Total} \\
\hline\\[-5pt]
$1s$ \\
 30 & -1.8191 & 47.4262 & -1.3817\,(7) & -58.7732\,(6) & -0.922\,(2) & -15.470\,(2) \\
 40 & -1.1463 & 23.4198 & -0.8725\,(2) & -29.1977\,(2) & -0.458\,(2) & -8.255\,(2) \\
 50 & -0.7824 & 13.4215\,(1) & -0.623\,(1) & -16.7579\,(1) & -0.261\,(1) & -5.003\,(2) \\
 60 & -0.5649 & 8.4658 & -0.480\,(1) & -10.5993\,(1) & -0.161\,(1) & -3.339\,(2) \\
 70 & -0.4242 & 5.7154 & -0.394\,(2) & -7.2105\,(1) & -0.105\,(1) & -2.418\,(2) \\
 83 & -0.3023 & 3.7005 & -0.322\,(3) & -4.7746 & -0.0654\,(8) & -1.764\,(3) \\
 92 & -0.2395 & 2.8545 & -0.293\,(2) & -3.7871 & -0.0499\,(7) & -1.515\,(2) \\
100 & -0.1903 & 2.3260 & -0.2775\,(4) & -3.2023 & -0.0422\,(3) & -1.3864\,(5) \\
\hline\\[-5pt]
$2s$ \\
 30 & -1.9540 & 87.4704\,(5) & -7.305\,(5) & -135.761\,(6) & -6.27\,(1) & -63.82\,(1) \\
 40 & -1.1141 & 44.4460\,(2) & -4.644\,(2) & -70.993\,(4) & -3.763\,(5) & -36.067\,(6) \\
 50 & -0.6395 & 26.3102\,(1) & -3.2972\,(7) & -42.716\,(1) & -2.478\,(6) & -22.820\,(6) \\
 60 & -0.3443 & 17.2011 & -2.5341\,(6) & -28.2078\,(7) & -1.732\,(5) & -15.617\,(5) \\
 70 & -0.1451 & 12.0764\,(1) & -2.0778\,(6) & -19.9526\,(4) & -1.272\,(1) & -11.371\,(1) \\
 83 & 0.0391 & 8.2685\,(1) & -1.755\,(2) & -13.8108\,(5) & -0.896\,(1) & -8.154\,(2) \\
 92 & 0.1443 & 6.6512 & -1.662\,(1) & -11.2475\,(5) & -0.727\,(1) & -6.842\,(2) \\
100 & 0.2370 & 5.6372 & -1.670\,(1) & -9.7065\,(5) & -0.620\,(2) & -6.122\,(2) \\
\hline\\[-5pt]
$2p_{1/2}$ \\
 30 & 0.0437 & 107.4798\,(4) & -3.696\,(7) & -144.371\,(5) & -7.137\,(7) & -47.68\,(1) \\
 40 & 0.0136 & 55.1287\,(1) & -2.439\,(3) & -76.167\,(2) & -4.275\,(2) & -27.739\,(4) \\
 50 & -0.0034 & 32.9318\,(1) & -1.7497\,(7) & -46.1202\,(3) & -2.827\,(2) & -17.768\,(2) \\
 60 & -0.0119 & 21.7304 & -1.3278\,(7) & -30.5892\,(2) & -1.994\,(2) & -12.192\,(2) \\
 70 & -0.0144 & 15.4027\,(1) & -1.0506\,(7) & -21.6861\,(1) & -1.466\,(3) & -8.814\,(3) \\
 83 & -0.0111 & 10.6814 & -0.8175\,(7) & -14.9849\,(2) & -1.015\,(3) & -6.147\,(3) \\
 92 & -0.0053 & 8.6696 & -0.7146\,(6) & -12.1248\,(2) & -0.793\,(2) & -4.968\,(2) \\
100 & 0.0027 & 7.4073 & -0.6578\,(9) & -10.3424\,(3) & -0.634\,(3) & -4.224\,(3) \\
\end{tabular}
\end{ruledtabular}
\end{table*}

\section{$\bm{F}$ term}

\begin{figure}
\centerline{
\resizebox{0.95\columnwidth}{!}{%
  \includegraphics{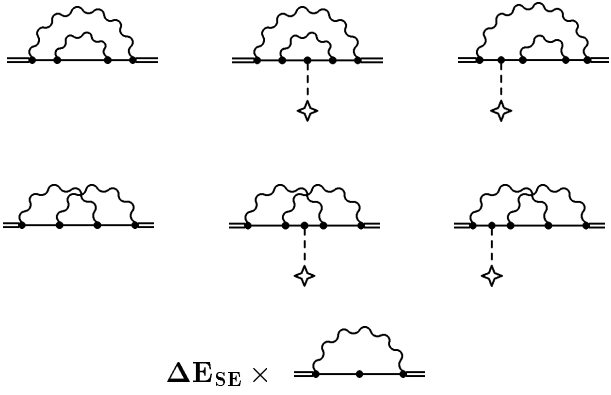}
}}
\vskip0.5cm \caption{Feynman diagrams contributing to the $F$ term. The single line denotes the
free electron, the double line denotes the bound electron,
the dashed line terminated by a cross denotes the interaction with the Coulomb field of the nucleus.
 \label{fig:fterm}}
\end{figure}

The $F$ term comprises a part of the UV subtraction terms introduced in the $M$ term, namely, those
that contain zero or one interaction with the binding Coulomb field in the electron propagators.
The corresponding Feynman diagrams are shown in Fig.~\ref{fig:fterm}. There are no IR divergences
in the $F$ term since the momenta of the initial and the final electron state are off mass-shell
and the virtuality $\rho = (m^2-p^2)/m^2 = (m^2-\vare_a^2 + \bfp^2)/m^2$ is strictly positive and
never vanishes.

It is natural to separate the $F$ term into the zero-potential $F$ part $\Delta E_{F,\, \rm zero}$
(comprising two-loop diagrams with no interactions with the Coulomb field in electron propagators),
the one-potential $F$ part $\Delta E_{F,\, \rm one}$ (comprising two-loop diagrams with one
interaction with the Coulomb field in electron propagators), and the reducible $F$ part $\Delta
E_{F,\, \rm red}$ (containing derivative of the one-loop diagram),
\begin{align}
\Delta E_{F} = \Delta E_{F,\, \rm zero} + \Delta E_{F,\, \rm one} + \Delta E_{F,\, \rm red}\,.
\end{align}

The zero-potential $F$ term can be written as
\begin{align}\label{eq:f0}
\Delta E_{F,\, \rm zero} = &\ \int \frac{d\bfp}{(2\pi)^3}\,
\psi^{\dag}_a(\bfp)\,\gamma^0\,
  \Sigma_{{\rm zero}, R}^{(2)}(p)\, \psi_a(\bfp)
 \nonumber \\
  \equiv &\ \lbr a| \gamma^0\Sigma_{{\rm zero}, R}^{(2)} |a\rbr \,,
\end{align}
where $p = (\vare_a,\bfp)$ is the 4-momentum of the bound electron and $\Sigma_{\rm zero,
R}^{(2)}(p)$ is the UV-finite part of the free two-loop self-energy operator $\Sigma_{\rm
zero}^{(2)}(p)$. The separation of UV divergences from the unrenormalized operator $\Sigma_{\rm
zero}^{(2)}(p)$ is performed by working in $D = 4-2\eps$ dimensions, expanding around $\eps = 0$
and identifying divergent $1/\eps$ and $1/\eps^2$ terms. It was demonstrated
\cite{yerokhin:03:epjd} that
\begin{align}
\Sigma_{\rm zero}^{(2)} - \delta m^{(2)} = (\cross{p}-m)\,B^{(2)}
   +\frac{\alpha C_{\eps}^2}{4\pi\eps}\, \Sigma_{R,4}^{(0)}(p)
     + \Sigma^{(2)}_{{\rm zero},R}(p)\,,
\end{align}
where $\delta m^{(2)}$ is the two-loop mass counterterm, $B^{(2)}$ is the two-loop renormalization
constant
\begin{align}
B^{(2)} = \frac{\alpha^2 C_{\eps}^2}{16\pi^2}
      \left(-\frac1{2\eps^2}+\frac3{4\eps} \right)\,,
\end{align}
$C_{\eps}$ is the two-loop prefactor
\begin{align} \label{fse2a}
 C_{\eps} = \Gamma(1+\eps)\,(4\pi)^{\eps}\,
   \left(\frac{\mu^2}{m^2}\right)^{\eps}\,,
\end{align}
and $\Sigma_{R,4}^{(0)}(p)$ is the UV-finite part of the free one-loop self-energy operator in
$D=4$ dimensions. The derivation of the UV-finite part of $\Sigma_{\rm zero, R}^{(2)}(p)$ is
discussed in detail in Ref.~\cite{yerokhin:03:epjd} and will not be repeated here.

The one-potential $F$ term is written as
\begin{align}
\Delta E_{F,\, \rm one} &\ =
\int \frac{d\bfp_1\, d\bfp_2}{(2\pi)^6}\, \psi^{\dag}_a(\bfp_1)\,\gamma^0\,
 V(\bfq)\,\Sigma_{{\rm one}, R}^{(2)}(p_1,p_2)\, \psi_a(\bfp_2) \nonumber \\
 & \equiv \lbr a| V\,\gamma^0\Sigma_{{\rm one}, R}^{(2)} |a\rbr\,,
\end{align}
where $p_1 = (\vare_a,\bfp_1)$ and $p_2 = (\vare_a,\bfp_2)$, $\bfq = \bfp_1-\bfp_2$, and
$\Sigma_{{\rm one}, R}^{(2)}$ is the UV-finite part of the two-loop vertex operator $\Sigma_{\rm
one}^{(2)}$. It was demonstrated \cite{yerokhin:03:epjd} that
\begin{align}
\Sigma^{(2)}_{\rm one}(p_1,p_2) =
  \gamma^0 L^{(2)} +
    \frac{\alpha C_{\eps}^2}{4\pi \eps}\,  \Gamma^0_{R,4}(p_1,p_2) +
             \Sigma^{(2)}_{{\rm one},R}(p_1,p_2)\,,
\end{align}
where $\Gamma^{\mu}_{R,4}(p_1,p_2)$ is the free one-loop vertex operator in $D=4$ dimensions, and
the two-loop renormalization constant $L^{(2)}$ is related to $B^{(2)}$ by the Ward identity,
$L^{(2)} = - B^{(2)}$. The derivation of the UV-finite two-loop vertex operator $\Sigma_{\rm one,
R}^{(2)}(p)$ is discussed in detail in Ref.~\cite{yerokhin:03:epjd} and will not be repeated here.

Finally, the reducible $F$ term can be expressed as \cite{yerokhin:03:epjd}
\begin{align}
\Delta E_{F,\, \rm red} &\ = \Delta E_{\rm SE} \, \lbr a|\gamma^0\,\frac{\partial}{\partial p^0}
\Sigma^{(0)}_{R,4}(p)\biggl|_{p^0 = \vare_a}|a\rbr
 \nonumber \\ &
 - \frac{\alpha}{4\pi}\,\lbr a|\gamma^0\,\frac{\partial}{\partial \eps}
    \frac{\Sigma^{(0)}_{R, D}(p)}{C_{\eps}}\biggr|_{\eps = 0} |a\rbr
 \nonumber \\ &
 - \frac{\alpha}{4\pi}\,\lbr a|\gamma^0\,V\, \frac{\partial}{\partial \eps}
    \frac{\Gamma^{0}_{R, D}(p_1,p_2)}{C_{\eps}}\biggr|_{\eps = 0} |a\rbr\,,
\end{align}
where $\Delta E_{\rm SE} $ is the one-loop self-energy correction to the energy. The second and the
third terms on the right-hand-side of the above equation contain matrix elements of the linear in
$\epsilon \equiv D-4$ parts of the one-loop operators $\Sigma^{(0)}_{R, D}$ and $\Gamma_{R,
D}(p_1,p_2)$ (see Eqs.~(248) and (256) of Ref.~\cite{yerokhin:03:epjd}), which yield finite
contributions when multiplied by divergent terms $\sim\!1/\epsilon$ from $\partial/(\partial p^0)
\Sigma^{(0)}_{D}(p)$.

Our numerical approach to the calculation of the $F$ term is described in
Ref.~\cite{yerokhin:03:epjd}. In the present work we use the same method, so it does not need to be
discussed here. The computation is relatively straightforward but time-consuming, in particular for
the one-potential contribution $\Delta E_{F,\, \rm one}$, which involves a 7-fold integration to be
performed numerically. As compared to our previous work \cite{yerokhin:03:epjd}, we adjusted all
numerical integrations, employing the extended Gauss-log quadratures \cite{pachucki:14:cpc}
alongside with the standard Gauss-Legendre quadratures, and enhanced the numerical accuracy of the
obtained results.

Our numerical results for the $F$ term for the $1s$, $2s$, and $2p_{1/2}$ states of hydrogen-like
ions with $Z = 30$-$100$ are presented in Table~\ref{tab:fterm}. The numerical accuracy of the
listed values is high enough so that it does not influence the total uncertainty of the final
results for the two-loop self-energy.

\begin{table*}
\caption{Numerical results for the $F$ term $\Delta E_{F}$, in units of $F(\Za)$  defined in Eq.~(\ref{FZa}). \label{tab:fterm}}
\begin{ruledtabular}
\begin{tabular}{ldddd}
 \multicolumn{1}{c}{$Z$}   & \multicolumn{1}{c}{$\Delta E_{F, \rm zero}$}  & \multicolumn{1}{c}{$\Delta E_{F, \rm one}$}  & \multicolumn{1}{c}{$\Delta E_{F, \rm red}$}  & \multicolumn{1}{c}{Total} \\
\hline\\[-5pt]
$1s$ \\
 30 & 24.66907\,(3) & -16.07999\,(8) & 36.13890\,(1) & 44.72798\,(9) \\
 40 & 11.61552\,(1) &  -9.24107\,(3) & 17.13164\,(1) & 19.50609\,(3) \\
 50 &  6.87772\,(1) &  -6.27573\,(2) &  9.42434\,(1) & 10.02633\,(2) \\
 60 &  4.66299\,(1) &  -4.64818\,(2) &  5.70902\,(1) &  5.72383\,(2) \\
 70 &  3.44582\,(1) &  -3.65961\,(1) &  3.71093\,(1) &  3.49714\,(1) \\
 83 &  2.54744\,(1) &  -2.92278\,(1) &  2.31342\,(1) &  1.93808\,(2) \\
 92 &  2.18316\,(2) &  -2.67665\,(1) &  1.76914\,(4) &  1.27566\,(5) \\
100 &  1.99368\,(8) &  -2.63549\,(3) &  1.46677\,(3) &  0.82496\,(9) \\
\hline\\[-5pt]
$2s$ \\
 30 & 90.7129\,(2)  & -50.90661\,(8) & 93.50484\,(2) & 133.3111\,(2) \\
 40 & 39.54636\,(4) & -23.2332\,(2)  & 47.42563\,(1) &  63.7388\,(2) \\
 50 & 21.25492\,(1) & -13.56096\,(7) & 28.03112\,(1) &  35.72508\,(8) \\
 60 & 13.17901\,(3) &  -9.28982\,(6) & 18.32256\,(2) &  22.21176\,(6) \\
 70 &  9.08349\,(1) &  -7.12317\,(5) & 12.89932\,(3) &  14.85964\,(6) \\
 83 &  6.34101\,(1) &  -5.76042\,(3) &  8.94329\,(3) &   9.52387\,(4) \\
 92 &  5.32624\,(3) &  -5.41592\,(1) &  7.33740\,(3) &   7.24773\,(5) \\
100 &  4.8297\,(2)  &  -5.49624\,(5) &  6.41643\,(5) &   5.7499\,(2) \\
\hline\\[-5pt]
$2p_{1/2}$ \\
 30 & 95.6913\,(2)  & -46.6496\,(3)  & 97.34783\,(4) & 146.3895\,(3) \\
 40 & 40.76330\,(3) & -20.2246\,(1)  & 48.18244\,(1) &  68.7211\,(2) \\
 50 & 21.48169\,(2) & -11.5546\,(1)  & 27.72745\,(1) &  37.65458\,(8) \\
 60 & 13.16105\,(3) &  -7.88637\,(5) & 17.60679\,(1) &  22.88147\,(6) \\
 70 &  9.05117\,(2) &  -6.02972\,(4) & 12.00600\,(2) &  15.02745\,(5) \\
 83 &  6.38819\,(1) &  -4.75684\,(3) &  7.93045\,(1) &   9.56180\,(3) \\
 92 &  5.43716\,(1) &  -4.29107\,(2) &  6.24689\,(1) &   7.39297\,(3) \\
100 &  4.98000\,(2) &  -4.11006\,(1) &  5.22996\,(1) &   6.09990\,(2) \\
%
%
\end{tabular}
\end{ruledtabular}
\end{table*}

\section{$\bm{P}$ term}

\begin{figure}[t]
\begin{center}\includegraphics[width=\columnwidth]{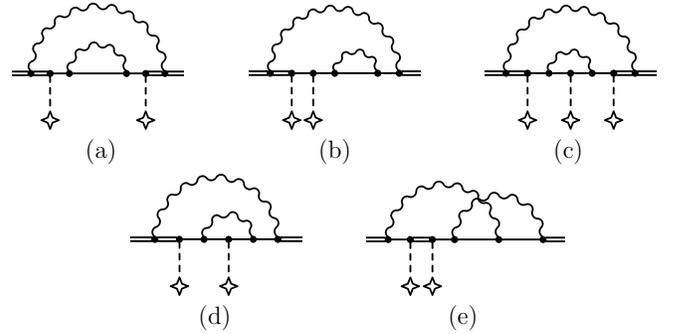}\end{center}%
\caption{\label{fig:pterm}
Feynman diagrams contributing to the $P$ term. }
\end{figure}

The $P$ term comprises the subtractions introduced in the $M$ term that contain two or more
interactions with the binding Coulomb field in the electron propagators. (We recall that the
subtractions with zero and one interactions with the Coulomb field are already accounted for by the
$F$ term.) The Feynman diagrams contributing to the $P$ term are shown in Fig.~\ref{fig:pterm}. The
distinct feature of these diagrams is that they contain {\em both} the bound-electron propagators
{\em and} the UV-divergent one-loop subgraphs (the free self-energy loop or the free vertex
subgraph). Since in our approach the isolation of UV divergences is performed in momentum space, we
have to evaluate the one-loop subgraphs in momentum space. The problem is that to treat the
bound-electron propagator ({\em i.e.}, the Dirac-Coulomb Green function) in momentum space as well
does not seem to be practically feasible.

Our solution \cite{yerokhin:01:sese} was to develop a method for computing the Dirac-Coulomb Green
function in the mixed coordinate-momentum representation ({\em i.e.}, one of the radial variables
in momentum space and the other radial variable in coordinate space). By now, we developed two
different numerical schemes for doing this. In our early works
\cite{yerokhin:01:sese,yerokhin:03:prl,yerokhin:05:sese,yerokhin:06:prl} we used the Fourier
transformation of the discrete representation of the spectrum of the Dirac-Coulomb Hamiltonian
obtained by the $B$-spline finite basis set method \cite{johnson:88,shabaev:04:DKB}. This approach
is convenient for practical implementations but is plagued by a slow convergence with respect to
the size of the basis set, which sets a limitation for the achievable numerical accuracy. In our
recent studies \cite{yerokhin:10:sese,yerokhin:09:sese} we developed a more effective solution of
this problem, which involves a numerical Fourier transformation of the analytical representation of
the Dirac-Coulomb Green function in terms of the Whittaker functions \cite{mohr:98}. This approach
provides the Dirac-Coulomb Green function in the mixed coordinate-momentum representation with a
high and controllable numerical precision, so that the limiting factor for the accuracy of the
final results becomes the convergence of the partial-wave expansion. In
Refs.~\cite{yerokhin:10:sese,yerokhin:09:sese} this approach was applied for the ground-state only.
In the present work we extend it to the excited states. This extension requires significant
modifications of the calculational scheme, which will be described next.

The contribution of Feynman diagrams shown in Fig.~\ref{fig:pterm} is conveniently represented as a
sum of three parts
\begin{align}
\Delta E_{P} =  \Delta E_{N,P,1} +  \Delta E_{N,P,2} + \Delta E_{O,P}\,,
\end{align}
which are discussed in turn below.

\begin{widetext}
\subsection{The first nested $\bm{P}$ contribution}

The first nested $P$ contribution is represented by two nested diagrams with the free self-energy
loop subgraph, shown in Fig.~\ref{fig:pterm}(a) and (b). The corresponding expression can be
written as \cite{yerokhin:10:sese}
\begin{eqnarray}\label{Peq2}
    \Delta E_{N,P,1} &=&
        \frac{\im}{2\pi}\int_{C_F} d\omega\,
          \Biggl\{
         \sum_{n_1n_2}
             \frac{
             \lbr an_2|I(\omega)|n_1a\rbr\,
                 \lbr n_1 V|\,{\cal S}(\vare_a-\omega)\,|V n_2\rbr }
        {(\vare_a-\omega-\vare_{n_1})(\vare_a-\omega-\vare_{n_2})}
  -
       \frac{
     \lbr a a''|I(\omega)|a'a\rbr\,
                 \lbr a' V|\,{\cal S}(\vare_a)\,|Va''\rbr}
                 {\omega^2}
        \nonumber \\
  &&{} + 2\,
         \sum_{n_1 \alpha _2 \alpha _3}
             \frac{
             \lbr a\alpha _2|I(\omega)|n_1a\rbr\, \lbr n_1|V|\alpha _3\rbr\,
                 \lbr \alpha _3 V|\,{\cal S}(\vare_a-\omega)\,|\alpha _2\rbr }
        {(\vare_a-\omega-\vare_{n_1})(\vare_a-\omega-\vare_{\alpha _3})}
                 \Biggr\}\,,
\end{eqnarray}
\end{widetext}
where the operator ${\cal S}$ is defined as
\begin{equation}\label{eq3}
{\cal S}(\vare) = \left. \frac1{\cross{p}-m}\,
\Sigma_R^{(0)}(p)\,\frac1{\cross{p}-m}\,\gamma^0 \right|_{p^0 = \vare}\,,
\end{equation}
and $\Sigma_R^{(0)}$ is renormalized free one-loop self-energy operator in $D = 4$ dimensions (see
Eq.~(\ref{eq1a})). The summations over $n_i$ are performed over the spectrum of the Dirac-Coulomb
Hamiltonian, the summations over $\alpha _i$ are performed over the spectrum of the {\em free}
Dirac Hamiltonian, and the states $a'$ and $a''$ denote the intermediate states that coincide with
the reference state $a$ (i.e., $\kappa_{a'} = \kappa_{a''} = \kappa_{a}$ and $n_{a'} = n_{a''} =
n_a$) except for the momentum projections ($\mu_{a'}$ and $\mu_{a''}$, correspondingly). Summations
over the magnetic substates (in particular, over $\mu_{a'}$ and $\mu_{a''}$) are implicit. In
Eq.~(\ref{Peq2}) we introduced the special notation $|V n\rbr$ for the product of the nuclear
Coulomb potential and the wave function. In coordinate space, it is just
\begin{equation}\label{Peq3a}
  |V n\rbr =   V(x)\,\psi_n(x)\,, \ \ \lbr n V| = \psi^{\dag}_n(x)\,V(x)\,.
\end{equation}
In momentum space, $|V n\rbr$ it is understood as a Fourier transform of the product
$V(x)\,\psi_n(x)$.  Matrix elements of ${\cal S}$ are assumed to be evaluated in momentum space,
whereas matrix operators of the electron-electron interaction operator $I(\omega)$ are assumed to
be evaluated in coordinate space.

The first term in the brackets in Eq.~(\ref{Peq2}) corresponds to the diagram in
Fig.~\ref{fig:pterm}(a), the second term is the IR subtraction that removes the reference-state IR
divergence in the first term, and the third term corresponds to the diagram in
Fig.~\ref{fig:pterm}(b). We note that the sum over $\alpha_2$ in the third term $\sum_{\alpha_2}
|\alpha_2\rbr \lbr \alpha_2|$ is the sum over the complete set of functions; it is inserted
artificially for the convenience of representation.

In order to make Eq.~(\ref{Peq2}) suitable for a numerical evaluation, we deform the contour of the
$\omega$ integration from $C_F$ to a new contour in the complex $\omega$ plane. In our previous
calculation \cite{yerokhin:10:sese}, we used the contour $C_{LH}$ described in
Appendix~\ref{app:se} with $\delta_y = 0$ (since only the ground state was considered in there). In
the present work we perform calculations for excited states. In this case there are virtual
intermediate states more deeply bound than the reference state. They would cause appearances of the
first-order ($\sim 1/(\delta-\omega)$) and second-order ($\sim 1/(\delta-\omega)^2$) singularities
on the low-energy part of the $C_{LH}$ contour with $\delta_y = 0$. In our evaluation of the $M$
term, we avoid such singularities by bending the low-energy part of the contour in the complex
plane (i.e., using $\delta_y > 0$). For the $P$ term, however, we found the contour with $\delta_y
> 0$  to be too difficult for a practical implementation. Instead, we prefer to use the contour
$C_{LH}$ with $\delta_y = 0$, treat the double-pole contributions separately, and handle the
first-order singularities by evaluating the principal value of the integral. Specifically, we
separate out the double-pole contributions $n_1 = n_2 = n$ with $0 < \vare_n < \vare_a$ and
calculate them separately, using the standard Wick rotation $\omega \to i\omega$ of the integration
contour.

\begin{widetext}
We thus write Eq.~(\ref{Peq2}) as a sum of two parts, $\Delta E_{N,P,1} = \Delta E_{N,P,1}^{\,a} +
\Delta E_{N,P,1}^{\,b}$, where the second part is the contribution of the $n_1 = n_2 = n$ states
with $0 < \vare_n < \vare_a$ and the first part is the remainder. The first term reads
\begin{eqnarray}\label{eq2a2}
    \Delta E_{N,P,1}^{\,a} &=&
        \frac{\im}{2\pi}\int_{C_{LH}}d\omega\,
          \Biggl\{
         \sum_{n_1n_2}
             \frac{
             \lbr an_2|I(\omega)|n_1a\rbr\,
                 \lbr n_1 V|\,{\cal S}(\vare_a-\omega)\,|V n_2\rbr }
        {(\vare_a-\omega-\vare_{n_1})(\vare_a-\omega-\vare_{n_2})}
-
  \frac{
     \lbr a a''|I(\omega)|a'a\rbr\,
                 \lbr a' V|\,{\cal S}(\vare_a)\,|Va''\rbr}
                 {\omega^2}
        \nonumber \\
  &&{} + 2\,
         \sum_{n_1 \alpha_2 \alpha_3}
             \frac{
             \lbr a\alpha_2|I(\omega)|n_1a\rbr\, \lbr n_1|V|\alpha_3\rbr\,
                 \lbr \alpha_3 V|\,{\cal S}(\vare_a-\omega)\,|\alpha_2\rbr }
        {(\vare_a-\omega-\vare_{n_1})(\vare_a-\omega-\vare_{\alpha_3})}
- \sum_{n}^{0 < \vare_n < \vare_a}
       \frac{
     \lbr an''|I(\omega)|n'a\rbr\,
                 \lbr n' V|\,{\cal S}(\vare_a-\omega)\,|Vn''\rbr}
                 {(\vare_a-\omega-\vare_n)^2}
                 \Biggr\}\,.\nonumber\\
\end{eqnarray}
In the above formula, the states $|n'\rbr$ and $|n''\rbr$ have the same energy $\vare_{n'} =
\vare_{n''} = \vare_n$ and quantum numbers $\kappa_{n'} = \kappa_{n''} = \kappa_n$ and $n' = n'' =
n$, but different values of the momentum projections, $\mu_{n'}$ and $\mu_{n''}$, respectively.
Summations over the magnetic substates (including $\mu_{n'}$ and $\mu_{n''}$) are implicit.  The
subtraction of the last term in the brackets of removes all {\em second-order} singularities on the
low-energy part of the contour. The remaining first-order singularities $(\sim 1/(\delta -
\omega))$ were handled by evaluating the principal value of the integral numerically (by using
integration quadratures symmetrical with respect to the position of the pole).

The second part $\Delta E_{N,P,1}^{\,b}$ is the contribution of the $n_1 = n_2 = n$ states with $0
< \vare_n <\vare_a$. It is transformed by applying the Wick rotation of the contour $\omega \to
i\omega$ and identifying the corresponding pole contributions. The result is
\begin{eqnarray}  \label{eq2b2}
    \Delta E_{N,P,1}^{b} &=&
\sum_{n}^{0<\vare_n< \vare_a} \Biggl[
     \lbr an''|I(\Delta_{an})|n'a\rbr\,
                \lbr n'V|{\cal S}^{\prime}(\vare_n) |Vn''\rbr
     -\lbr an''|I{}^{\prime}(\Delta_{an})|n'a\rbr\,
                \lbr n'V|{\cal S}(\vare_n) |Vn''\rbr
 \nonumber \\ &&
        -
        \frac{1}{\pi}\,{\rm Re}\int_{0}^{\infty}d\omega\,
             \frac{
             \lbr an''|I(\im\omega)|n'a\rbr\,
                 \lbr n' V|\,{\cal S}(\vare_a-\im \omega)\,|V n''\rbr }
        {(\Delta_{an}-\im \omega)^2}
        \Biggr]
                 \,,
\end{eqnarray}
where $\Delta_{an} = \vare_a - \vare_n$ and $I'$ and ${\cal S}'$ denote the derivatives over the
energy argument.

In order to evaluate Eqs.~(\ref{eq2a2}) and (\ref{eq2b2}) numerically, we rewrite all sums over the
Dirac spectrum in terms of the Green functions. In particular, for the first term in the brackets
in Eq.~(\ref{eq2a2}) we use the representation
\begin{align} \label{eq4a1}
         \sum_{n_1n_2}
             \frac{
                | n_1\rbr
                 \lbr n_1 V|\,{\cal S}(E)\,|V n_2\rbr \lbr n_2|}
        {(E-\vare_{n_1})(E-\vare_{n_2})} =
        \int \frac{d\bfp}{(2\pi)^3}\, G_V(E,\bfx_1,\bfp)\,
{\cal S}(E,\bfp)\,
         G_V(E,\bfp,\bfx_2)\,,
\end{align}
where $G_V$ denotes the (Fourier transform of the) product $G V$,
\begin{equation} \label{eq3a1}
G_V(\vare,\bfx_1,\bfp) =
        \int d\bfx_2 \, e^{i\bfp\cdot \bfx_2} \,
       G(\vare,\bfx_1,\bfx_2) \, V(\bfx_2)
         \,,
\end{equation}
\begin{equation} \label{eq3a2}
G_V(\vare,\bfp,\bfx_2) =
        \int d\bfx_1 \, e^{-i\bfp\cdot \bfx_1}\,
    V(\bfx_1)\,
          G(\vare,\bfx_1,\bfx_2) \, .
\end{equation}
For the third term in the brackets in Eq.~(\ref{eq2a2}) we use the following representation
\begin{align} \label{eq4a2}
         \sum_{n_1 \alpha_2 \alpha_3}
             \frac{
                | n_1\rbr
                 \lbr n_1|V|\alpha_3\rbr\,
                 \lbr \alpha_3 V|\,{\cal S}(E)\,|\alpha_2\rbr \lbr \alpha_2|}
        {(E-\vare_{n_1})(E-\vare_{\alpha_3})} =
        \int \frac{d\bfp}{(2\pi)^3}\, G^{(1+)}_V(E,\bfx_1,\bfp)\,
  \frac1{\gamma^0E -\bgamma\cdot\bfp-m}\,
\Sigma_R^{(0)}(E,\bfp)\,
         G^{(0)}(E,\bfp,\bfx_2)\,,
\end{align}
where $G^{(1+)}_V$ is the part of $G_V$ with one and more Coulomb interactions, $G_V \equiv
G^{(0)}_V + G^{(1+)}_V$.

The main difficulty of the numerical evaluation of the $P$ term is that the computation of the
Fourier transform of the Dirac-Coulomb Green function is rather time-consuming (since it is done by
evaluating the momentum integration numerically, see Appendix A of Ref.~\cite{yerokhin:10:sese} for
details). The key idea is to perform the radial integrations over $x_1$ and $x_2$ {\em before} the
integration over $p$. For a given value of $p$, we compute and store the Fourier transform of the
Dirac-Coulomb Green function for all points of the radial $x$ grid that are needed for computation
of the radial integrals. The scheme is described in detail in Ref.~\cite{yerokhin:10:sese}.

A difficulty arises in the numerical evaluation of Eq.~(\ref{eq4a2}), due to the presence of the
free Green function $G^{(0)}(E,\bfp,\bfx_2)$ in the mixed momentum-coordinate representation. For
large values of $p$, $G^{(0)}(E,\bfp,\bfx_2)$ is a strongly oscillating function of $x_2$. Rapid
oscillations cause the radial integration over $x_2$ to converge slowly and require very dense
integration grids. We address this difficulty by noting that $G^{(0)}(E,\bfp,\bfx_2)$ can be
expressed analytically in terms of the spherical Bessel functions (see Appendix B of
Ref.~\cite{yerokhin:10:sese}) and thus the integral over $x_2$ is essentially the Bessel transform
of a relatively simple function, which can be computed by the same method as the Fourier transform
of the Dirac-Coulomb Green function. By using this approach, we were able to achieve a very good
stability of the radial integrations in our computations.

\subsection{The second nested $\bm{P}$ contribution}

The second nested $P$ contribution is represented by nested diagrams containing the free vertex
subgraph. They are shown in Fig.~\ref{fig:pterm}(c) and (d). The corresponding expression can be
written as \cite{yerokhin:10:sese}
\begin{eqnarray}\label{Peq4}
    \Delta E_{N,P,2} &=&
        \frac{\im}{2\pi}\int_{C_F}d\omega\,
          \Biggl\{
         \sum_{n_1n_2}
             \frac{
             \lbr an_2|I(\omega)|n_1a\rbr\,
                 \lbr n_1 V|\,{\cal G}(\vare_a-\omega)\,|V n_2\rbr }
        {(\vare_a-\omega-\vare_{n_1})(\vare_a-\omega-\vare_{n_2})}
  -
       \frac{
     \lbr aa''|I(\omega)|a'a\rbr\,
                 \lbr a' V|\,{\cal G}(\vare_a)\,|Va''\rbr}
                 {\omega^2}
        \nonumber \\
  &&{} + 2\,
         \sum_{n_1 \alpha_2}
             \frac{
             \lbr a\alpha_2|I(\omega)|n_1a\rbr\,
                 \lbr n_1 V|\,{\cal G}(\vare_a-\omega)\,|\alpha_2\rbr }
        {\vare_a-\omega-\vare_{n_1}}
                 \Biggr\}\,,
\end{eqnarray}
where the operator ${\cal G}$ is defined as
\begin{equation}\label{eqI5}
{\cal G}(\vare,\bfp_1,\bfp_2) = \left. \frac1{\cross{p}_1-m}\, V(\bfq)\,
\Gamma_R^{\,0}(p_1,p_2)\,\frac1{\cross{p}_2-m}\,\gamma^0 \right|_{p_1^0 = p_2^0 =
\vare}\,,
\end{equation}
$\bfq = \bfp_1-\bfp_2$, $V$ is the Coulomb potential, and $\Gamma_R^{\,0}$ is the time component of
the renormalized free vertex operator in $D = 4$ dimensions (see Eq.~(\ref{eq:onepot})).

In principle, we could have evaluated Eq.~(\ref{Peq4}) in full analogy with the scheme described
for the first nested $P$ term. Such evaluation, however, would be much more time consuming, since
matrix elements of the operator ${\cal G}$ involve two momentum integrations (over $\bfp_1$ and
$\bfp_2$) and are complicated by the (integrable) Coulomb singularity at $\bfq = 0$. In order to
compute the momentum integrations efficiently, it is desirable to separate out the Coulomb
singularity. We achieve this by subtracting and re-adding the vertex operator with the zero
transferred momentum and performing one of the momentum integration in the separated contribution
analytically. Specifically, we split the vertex operator into the regular ($r$) and irregular ($i$)
parts with help of Ward identity as
\begin{align} \label{eqI5b}
\Gamma_R^{\,0}(p_1,p_2) =
\Bigl[ \Gamma_{\rm R}^{\,0}(p_1,p_2)- \frac12\,\Gamma_{\rm R}^{\,0}(p_1,p_1)- \frac12\,\Gamma_{\rm
R}^{\,0}(p_2,p_2)\Bigr] + \Bigl[
- \frac12\,\Sigma_R^{(0)^{\prime}}\!(p_1)
    - \frac12\,\Sigma_R^{(0)^{\prime}}\!(p_2) \Bigr] \equiv
\Gamma_{r}^{\,0}(p_1,p_2) + \Gamma_{i}^{\,0}(p_1,p_2)
    \,,
\end{align}
where the prime denotes derivative with respect to $p^0$. The regular operator $\Gamma_{
r}^{\,0}(p_1,p_2)$ vanishes at $\bfp_1 = \bfp_2$, thus removing the Coulomb singularity in the
corresponding integral. In the irregular term, the Coulomb singularity is integrated out
analytically, by using the the Dirac equation and the definition of the Green function. In
particular, we make use of the identity
\begin{align}
\int \frac{d\bfp_2}{(2\pi)^3}\, &\, V(\bfq)\,
     \frac1{\gamma^0E-\bgamma\cdot\bfp_2-m}\, G_V(E,\bfp_2,\bfx_2) =
 G_V^{(1+)}(E,\bfp_1,\bfx_2)\,.
\end{align}

The next step is to deform the integration contour of the $\omega$ integration, in full analogy
with the procedure described for the first nested $P$ contribution, which splits the regular and
the irregular terms into the $a$ and $b$ parts. In the result, the second nested $P$ contribution
is written as a sum of four terms,
\begin{align} \label{eqI6}
    \Delta E_{N,P,2} = \Delta E_{N,P,2}^{r,a} + \Delta E_{N,P,2}^{r,b} + \Delta E_{N,P,2}^{i,a} + \Delta E_{N,P,2}^{i,b}\,,
\end{align}
where
\begin{eqnarray}\label{eqI6a}
    \Delta E_{N,P,2}^{r,a} &=&
        \frac{\im}{2\pi}\int_{C_{LH}}d\omega\,
          \Biggl\{
         \sum_{n_1n_2}
             \frac{
             \lbr an_2|I(\omega)|n_1a\rbr\,
                 \lbr n_1 V|\,{\cal G}_{r}(\vare_a-\omega)\,|V n_2\rbr }
        {(\vare_a-\omega-\vare_{n_1})(\vare_a-\omega-\vare_{n_2})}
  -        \frac{
     \lbr aa''|I(\omega)|a'a\rbr\,
                 \lbr a' V|\,{\cal G}_{r}(\vare_a)\,|Va''\rbr}
                 {\omega^2}
        \nonumber \\
  &&{} + 2\,
         \sum_{n_1 \alpha_2}
             \frac{
             \lbr a\alpha_2|I(\omega)|n_1a\rbr\,
                 \lbr n_1 V|\,{\cal G}_{r}(\vare_a-\omega)\,|\alpha_2\rbr }
        {\vare_a-\omega-\vare_{n_1}}
  - \sum_n^{0 < \vare_n < \vare_a}
       \frac{
     \lbr an''|I(\omega)|n'a\rbr\,
                 \lbr n' V|\,{\cal G}_{r}(\vare_a-\omega)\,|Vn''\rbr}
                 {(\vare_a-\omega-\vare_n)^2}
                 \Biggr\}\,,
                 \nonumber \\
\end{eqnarray}
\begin{eqnarray}  \label{eqI6b}
    \Delta E_{N,P,2}^{r,b} &=&
\sum_n^{0<\vare_n< \vare_a} \Biggl[
     \lbr an''|I(\Delta_{an})|na'\rbr\,
                \lbr n'V|{\cal G}_{r}^{\prime}(\vare_n) |Vn''\rbr
     -\lbr an''|I{}^{\prime}(\Delta_{an})|n'a\rbr\,
                \lbr n'V|{\cal G}_{r}(\vare_n) |Vn''\rbr
 \nonumber \\ &&
        -
        \frac{1}{\pi}\,{\rm Re}\int_{0}^{\infty}d\omega\,
             \frac{
             \lbr an''|I(\im\omega)|n'a\rbr\,
                 \lbr n' V|\,{\cal G}_{r}(\vare_a-\im \omega)\,|V n''\rbr }
        {(\vare_a-\im \omega-\vare_{n})^2}
        \Biggr]
                 \,,
\end{eqnarray}
\begin{eqnarray}\label{eqI6c}
    \Delta E_{N,P,2}^{i,a} &=&
        -\frac{\im}{2\pi}\int_{C_{LH}}d\omega\,
          \Biggl\{
         \sum_{n_1\alpha_1 n_2}
             \frac{
             \lbr an_2|I(\omega)|n_1a\rbr\,
                 \lbr n_1| V|\alpha_1\rbr \lbr \alpha_1|\,\Sigma^{(0)^{\prime}}_R\!(\vare_a-\omega)\,|V n_2\rbr }
        {(\vare_a-\omega-\vare_{n_1})(\vare_a-\omega-\vare_{\alpha_1})(\vare_a-\omega-\vare_{n_2})}
  -        \frac{
     \lbr aa''|I(\omega)|a'a\rbr\,
                 \lbr a'|\,\Sigma^{(0)^{\prime}}_R\!(\vare_a)\,|Va''\rbr}
                 {\omega^2}
        \nonumber \\
  &&{} +
  \sum_{n_1\alpha_1\alpha_2}
             \frac{
             \lbr a\alpha_2|I(\omega)|n_1a\rbr\,
                 \lbr n_1| V|\alpha_1\rbr \lbr \alpha_1 V|\,\Sigma^{(0)^{\prime}}_R\!(\vare_a-\omega)\,|\alpha_2\rbr }
        {(\vare_a-\omega-\vare_{n_1})(\vare_a-\omega-\vare_{\alpha_1})(\vare_a-\omega-\vare_{\alpha_2})}
  - \sum_n^{0 < \vare_n < \vare_a}
       \frac{
     \lbr an''|I(\omega)|n'a\rbr\,
                 \lbr n'|\,\Sigma^{(0)^{\prime}}_R\!(\vare_a-\omega)\,|Vn''\rbr}
                 {(\vare_a-\omega-\vare_n)^2}
                 \Biggr\}\,,
                 \nonumber \\
\end{eqnarray}
\begin{eqnarray}  \label{eqI6d}
    \Delta E_{N,P,2}^{i,b} &=&
(-1)\sum_{n}^{0<\vare_n< \vare_a} \Biggl[
     \lbr an''|I(\Delta_{an})|n'a\rbr\,
                \lbr n'|\Sigma^{(0)^{\prime\prime}}_R\!(\vare_n) |Vn''\rbr
     -\lbr an''|I{}^{\prime}(\Delta_{an})|n'a\rbr\,
                \lbr n'|\Sigma^{(0)^{\prime}}_R\!(\vare_n) |Vn''\rbr
 \nonumber \\ &&
        -
        \frac{1}{\pi}\,{\rm Re}\int_{0}^{\infty}d\omega\,
             \frac{
             \lbr an''|I(\im\omega)|n'a\rbr\,
                 \lbr n' |\,\Sigma^{(0)^{\prime}}_R\!(\vare_a-\im \omega)\,|V n''\rbr }
        {(\vare_a-\im \omega-\vare_{n})^2}
        \Biggr]
                 \,,
\end{eqnarray}
where ${\cal G}_{r}(\vare)$ is obtained from Eq.~(\ref{eqI5}) by the substitution
$\Gamma_R^{0}(p_1,p_2) \to \Gamma_{r}^{0}(p_1,p_2)$. Out of the four terms in the right-hand-side
of Eq.~(\ref{eqI6}), only the two first ones contain two momentum integrations. They involve the
operator ${\cal G}_{r}$, which is regular at $\bfp_1 = \bfp_2$. The last two terms in
Eq.~(\ref{eqI6}) contain one momentum integration, and their evaluation is similar to that for the
first nested contribution.

For the numerical evaluation, we rewrite Eqs.~(\ref{eqI6a})-(\ref{eqI6d}) in terms of the Green
function, identifying, in particular,
\begin{align} \label{eqI6b}
         \sum_{n_1}
             \frac{| n_1\rbr  \lbr n_1 V|}
        {E-\vare_{n_1}} =
        G_V(E,\bfx_1,\bfp_1)\,,
\end{align}
\begin{align} \label{eqI7}
         \sum_{n_1\alpha_1}
             \frac{| n_1\rbr
                 \lbr n_1| V |\alpha_1\rbr\lbr \alpha_1 V|}
        {(E-\vare_{n_1})(E-\vare_{\alpha_1})} =
        G_V^{(1+)}(E,\bfx_1,\bfp_1)\,.
\end{align}

We checked that the expressions with and without the separation (\ref{eqI5b}) give the results
consistent with each other within the estimated numerical error.

\subsection{Overlapping $\bm P$ term}

The overlapping $P$ term is represented by Feynman diagram in Fig.~\ref{fig:pterm}(e). The
corresponding expression is written as \cite{yerokhin:10:sese}
\begin{align} \label{eqO1}
\Delta E_{O,P}  = -4i\alpha \int_{C_F} d\omega\,
\int \frac{d\bfp_1}{(2\pi)^3}\,
        \frac{d\bfp_2}{(2\pi)^3}\,
        \int d\bfz \,   &\
                \frac{\exp(-i\bfq\cdot \bfz)}{\omega^2 -\bfq^2+ i0}\,
\psi_a^{\dag}(\bfz)\, \alpha_{\mu}\,
                \nonumber \\ & \times
 G_V^{(1+)}(\vare_a-\omega,\bfz,\bfp_1)\,
        \widetilde{\Gamma}^{\mu}(\vare_a-\omega,\bfp_1;\vare_a,\bfp_2)\,
                        \psi_a(\bfp_2)
\,,
\end{align}
where
$$
\widetilde{\Gamma}^{\mu}(E,\bfp_1;\vare_a,\bfp_2) \equiv \frac1{\gamma^0 E-\bgamma\cdot\bfp_1-m}\,\,
        \Gamma_R^{\mu}(E,\bfp_1;\vare_a,\bfp_2)\,,
$$
and $\Gamma_R^{\mu}$ is the renormalized one-loop free vertex operator in $D = 4$ dimensions (see
Eq.~(257) of Ref.~\cite{yerokhin:03:epjd}). In our previous investigations
\cite{yerokhin:03:epjd,yerokhin:10:sese} we evaluated Eq.~(\ref{eqO1}) by performing the Wick
rotation of the $\omega$ integration contour and evaluating separately the pole contributions. For
the excited states and low values of the nuclear charges considered in the present work, however,
the behaviour of the integrand becomes rather complicated for small $\omega$, aquiring a rapidly
changing structure which is difficult to integrate numerically. In order to make the integrand
behave more smoothly in the region of small $\omega$, we subtract and re-add the following
contribution
\begin{align}  \label{eqO2}
\Delta E_{O,P}^{\rm \, subtr}  = -4i\alpha \int_{C_F} d\omega\,
\int \frac{d\bfp_1}{(2\pi)^3}\,
        \frac{d\bfp_2}{(2\pi)^3}\,
        \int d\bfz \,   &\
                \frac{\exp(-i\bfq\cdot \bfz)}{\omega^2 -\bfq^2+ i0}\,
\psi_a^{\dag}(\bfz)\, \alpha_{\mu}\,
                \nonumber \\ & \times \sum_{k}\,
 \frac{\psi_k(\bfz)\,(\psi_{k}V)^{\dag}(\bfp_1)}{\vare_a-\omega-\vare_k}\,\,
        \widetilde{\Gamma}^{\mu}(\vare_a,\bfp_1;\vare_a,\bfp_2)\,
                        \psi_a(\bfp_2)
\,,
\end{align}
where the index $k$ run over the virtual bound states with energies $0 < \vare_k < \vare_a +
\delta$, where $\delta$ is some (reasonably small) positive parameter. It can be immediately seen
that we obtained $\Delta E_{O,P}^{\rm \, subtr}$ from $\Delta E_{O,P}$ by neglecting $\omega$ in
the vertex operator and retaining only the lowest-lying virtual bound states in the spectral
decomposition of the electron propagator. The summation over $k$ in Eq.~(\ref{eqO2}) thus includes
virtual states more deeply bound than the reference state and in addition, virtual states that are
close in energy to the reference state. The subtraction of $\Delta E_{O,P}^{\rm \, subtr}$ removes
the complicated structure of the integrand for small $\omega$. On the other hand, the fact that the
vertex operator $\widetilde{\Gamma}$ in Eq.~(\ref{eqO2}) does not depend on the energy of the
virtual photon $\omega$ allows us to perform the integral over $\omega$ analytically by the Cauchy
theorem.

We thus write
$$
\Delta E_{O,P} = \left(\Delta E_{O,P} - \Delta E_{O,P}^{\rm\, subtr}\right) + \Delta E_{O,P}^{\rm\, subtr}\,.
$$
In the first part we make the Wick rotation of the $\omega$ integration contour $\omega \to
i\omega$. As a result, this part is separated into the pole term $\Delta E_{O,P}^{\rm\, pole}$ and
the integral over the imaginary axis, $\Delta E_{O,P}^{\rm\, Im}$. In the second part we perform
the $\omega$ integration analytically by Cauchy theorem. The result is
\begin{align}  \label{eqO3}
\Delta E_{O,P}  = \Delta E_{O,P}^{\rm \, pole} + \Delta E_{O,P}^{\rm\, Im} + \Delta E_{O,P}^{\rm\, subtr}
\,,
\end{align}
where
\begin{align}  \label{eqO4}
\Delta E_{O,P}^{\,\rm pole}  = -8\pi\alpha \sum_{0 < \vare_n < \vare_a}
\int \frac{d\bfp_1}{(2\pi)^3}\,
        \frac{d\bfp_2}{(2\pi)^3}\,
        \int d\bfz \,   &\
                \frac{e^{-i\bfq\cdot \bfz}}{(\vare_a-\vare_n)^2 -\bfq^2+ i0}\,
\psi_a^{\dag}(\bfz)\, \alpha_{\mu}\,
                \nonumber \\ & \times
 \psi_n(\bfz)\,(\psi_{n}V)^{\dag}(\bfp_1)\,
        \bigl[ \widetilde{\Gamma}^{\mu}(\vare_n,\bfp_1;\vare_a,\bfp_2)
        - \widetilde{\Gamma}^{\mu}(\vare_a,\bfp_1;\vare_a,\bfp_2)\bigr]\,
                        \psi_a(\bfp_2)
\,,
\end{align}
\begin{align}  \label{eqO5}
\Delta E_{O,P}^{\,\rm Im} &\ = 8\alpha \int_0^{\infty}d\omega\,
\int \frac{d\bfp_1}{(2\pi)^3}\,
        \frac{d\bfp_2}{(2\pi)^3}\,
        \int d\bfz \,
                \frac{e^{-i\bfq\cdot \bfz}}{-\omega^2 -\bfq^2}\,
\psi_a^{\dag}(\bfz)\, \alpha_{\mu}\,
                \nonumber \\ & \times
\biggl[
 G_V^{(1+)}(\vare_a-i\omega,\bfz,\bfp_1)\,
         \widetilde{\Gamma}^{\mu}(\vare_a-i\omega,\bfp_1;\vare_a,\bfp_2)
- \sum_k \frac{\psi_k(\bfz)\,(\psi_{k}V)^{\dag}(\bfp_1)}{\vare_a-i\omega-\vare_k}\,\,
         \widetilde{\Gamma}^{\mu}(\vare_a,\bfp_1;\vare_a,\bfp_2)\biggr]\,
                        \psi_a(\bfp_2)
\,,
\end{align}
\begin{align}  \label{eqO6}
\Delta E_{O,P}^{\,\rm subtr}  = -4\pi\alpha \sum_{k}
\int \frac{d\bfp_1}{(2\pi)^3}\,
        \frac{d\bfp_2}{(2\pi)^3}\,
        \int d\bfz \,   &\
                \frac{e^{-i\bfq\cdot \bfz}}{q(\vare_a-q-\vare_k )}\,
\psi_a^{\dag}(\bfz)\, \alpha_{\mu}\,
 \psi_k(\bfz)\,(\psi_{k}V)^{\dag}(\bfp_1)\,
        \widetilde{\Gamma}^{\mu}(\vare_a,\bfp_1;\vare_a,\bfp_2)\,
                        \psi_a(\bfp_2)
\,.
\end{align}
\end{widetext}
The number of states included into the summation over $k$ can be varied; it also serves as a
cross-check of the correctness of the numerical procedure. For the ground state we performed
calculations with and without the separation of $\Delta E_{O,P}^{\,\rm subtr}$; the results were
found to be in perfect agreement with each other.

The numerical procedure for evaluation of the $P$ term was described for the $1s$ state in
Ref.~\cite{yerokhin:10:sese}. This procedure can be directly generalized to the case of excited
states considered here. Our numerical results for the $P$ term are presented in
Table~\ref{tab:pterm}. The dominant numerical uncertainty of the listed results comes from the
truncation of the partial-wave expansion and extrapolation of the expansion tail.

\begin{table*}
\caption{Numerical results for the $P$ term $\Delta E_{P}$, in units of $F(\Za)$  defined in Eq.~(\ref{FZa}). \label{tab:pterm}}
\begin{ruledtabular}
\begin{tabular}{ldddd}
 \multicolumn{1}{c}{$Z$}   & \multicolumn{1}{c}{$\Delta E_{N,P,1}$}  & \multicolumn{1}{c}{$\Delta E_{N,P,2}$}
 & \multicolumn{1}{c}{$\Delta E_{O,P}$}  & \multicolumn{1}{c}{Total} \\
\hline\\[-5pt]
$1s$ \\
 30 & -23.8156\,(8) & 44.810\,(2) & -50.408\,(2) & -29.413\,(2) \\
 40 & -9.0139\,(4) & 17.6060\,(8) & -20.1711\,(8) & -11.579\,(1) \\
 50 & -4.3392\,(3) & 8.4018\,(6) & -9.5503\,(6) & -5.4877\,(9) \\
 60 & -2.4456\,(2) & 4.5451\,(4) & -5.0659\,(4) & -2.9664\,(6) \\
 70 & -1.5202\,(1) & 2.6715\,(2) & -2.9353\,(3) & -1.7841\,(4) \\
 83 & -0.8653\,(1) & 1.4268\,(1) & -1.6306\,(3) & -1.0690\,(4) \\
 92 & -0.5542\,(1) & 0.9091\,(1) & -1.1901\,(3) & -0.8352\,(3) \\
100 & -0.2985\,(5) & 0.5426\,(1) & -0.9791\,(3) & -0.7349\,(6) \\
\hline\\[-5pt]
$2s$ \\
 30 & -27.003\,(9) & 117.320\,(8) & -160.260\,(8) & -69.94\,(1) \\
 40 & -3.711\,(4) & 46.364\,(6) & -71.083\,(4) & -28.430\,(9) \\
 50 & 1.318\,(2) & 21.768\,(4) & -37.036\,(3) & -13.949\,(6) \\
 60 & 2.285\,(1) & 11.235\,(4) & -21.451\,(2) & -7.932\,(5) \\
 70 & 2.276\,(1) & 5.980\,(3) & -13.432\,(2) & -5.176\,(4) \\
 83 & 2.085\,(1) & 2.291\,(2) & -8.043\,(2) & -3.668\,(3) \\
 92 & 2.1074\,(8) & 0.563\,(2) & -6.002\,(2) & -3.332\,(3) \\
100 & 2.3637\,(8) & -0.868\,(1) & -4.870\,(2) & -3.375\,(2) \\
\hline\\[-5pt]
$2p_{1/2}$ \\
 30 & -50.451\,(6) & 142.357\,(7) & -190.46\,(1) & -98.56\,(1) \\
 40 & -11.943\,(3) & 56.512\,(6) & -85.371\,(6) & -40.802\,(9) \\
 50 & -1.854\,(2) & 26.920\,(6) & -44.726\,(4) & -19.659\,(7) \\
 60 & 1.063\,(1) & 14.433\,(4) & -25.902\,(3) & -10.405\,(5) \\
 70 & 1.827\,(1) & 8.397\,(3) & -16.091\,(2) & -5.867\,(4) \\
 83 & 1.8854\,(8) & 4.482\,(2) & -9.359\,(2) & -2.992\,(3) \\
 92 & 1.7902\,(6) & 2.941\,(2) & -6.704\,(2) & -1.972\,(3) \\
100 & 1.7531\,(7) & 1.933\,(1) & -5.123\,(2) & -1.437\,(2) \\
\end{tabular}
\end{ruledtabular}
\end{table*}

\section{Results and discussion}

\begin{table*}
\caption{The two-loop self-energy correction, in terms of $F(\Za)$ defined in Eq.~(\ref{FZa}). $G_{\rm h.o.}$ is the
higher-order remainder function defined by Eq.~(\ref{Gho}). \label{tab:sese}}
\begin{ruledtabular}
\begin{tabular}{lddddddd}
 \multicolumn{1}{c}{$Z$}   & \multicolumn{1}{c}{$\Delta E_{\rm LAL}$}
    & \multicolumn{1}{c}{$\Delta E_{F}$}
        & \multicolumn{1}{c}{$\Delta E_{P}$}
            & \multicolumn{1}{c}{$\Delta E_{M}$}
                & \multicolumn{1}{c}{Total}
                & \multicolumn{1}{c}{Previous \cite{yerokhin:05:sese,yerokhin:06:prl}}
                    & \multicolumn{1}{c}{$G_{\rm h.o.}$}       \\
\hline\\[-5pt]
$1s$ \\
 30 & -0.7565 & 44.7280\,(1)  & -29.413\,(2) & -15.470\,(2) & -0.912\,(3)  & -0.90\,(3) & -70.39\,(7) \\
 40 & -0.8711 & 19.5061       & -11.579\,(1) & -8.255\,(2)  & -1.199\,(3)  & -1.19\,(3) & -58.35\,(3) \\
 50 & -0.9734 & 10.0263       & -5.4877\,(9) & -5.003\,(2)  & -1.438\,(2)  & -1.44\,(3) & -47.42\,(2) \\
 60 & -1.0825 & 5.7238        & -2.9664\,(6) & -3.339\,(2)  & -1.664\,(2)  & -1.67\,(2) & -37.460\,(8) \\
 70 & -1.2161 & 3.4971        & -1.7841\,(4) & -2.418\,(2)  & -1.922\,(2)  & -1.89\,(3) & -28.399\,(8) \\
 83 & -1.4658 & 1.9381        & -1.0690\,(4) & -1.764\,(3)  & -2.361\,(3)  & -2.35\,(1) & -17.798\,(9) \\
 92 & -1.7342 & 1.2757        & -0.8352\,(3) & -1.515\,(2)  & -2.809\,(2)  & -2.78\,(1) & -11.220\,(6) \\
100 & -2.0990 & 0.8250\,(1)   & -0.7349\,(6) & -1.3864\,(5) & -3.3953\,(8) & -3.381\,(8)& -5.934\,(2) \\
\hline\\[-5pt]
$2s$ \\
 30 & -0.4650\,(1) & 133.3111\,(2) & -69.94\,(1)  & -63.82\,(1)  & -0.91\,(2)  && -57.9\,(4) \\
 40 & -0.5155 & 63.7388\,(2)       & -28.430\,(9) & -36.067\,(6) & -1.27\,(1)  && -47.5\,(1) \\
 50 & -0.5695 & 35.7251\,(1)       & -13.949\,(6) & -22.820\,(6) & -1.613\,(8) && -38.19\,(6) \\
 60 & -0.6434 & 22.2118\,(1)       & -7.932\,(5)  & -15.617\,(5) & -1.980\,(7) & -1.98\,(7) & -29.79\,(4) \\
 70 & -0.7539 & 14.8596\,(1)       & -5.176\,(4)  & -11.371\,(1) & -2.442\,(4) & -2.45\,(6) & -22.36\,(2) \\
 83 & -0.9956 & 9.5239             & -3.668\,(3)  & -8.154\,(2)  & -3.294\,(4) & -3.30\,(4) & -13.97\,(1) \\
 92 & -1.2839 & 7.2477             & -3.332\,(3)  & -6.842\,(2)  & -4.209\,(3) & -4.22\,(3) & -9.086\,(7) \\
100 & -1.7071 & 5.7499\,(2)        & -3.375\,(2)  & -6.122\,(2)  & -5.454\,(3) & -5.46\,(7) & -5.539\,(6) \\
\hline\\[-5pt]
$2p_{1/2}$ \\
 30 & 0.0255 & 146.3895\,(3) & -98.56\,(1)  & -47.68\,(1)  & 0.18\,(2)   && -0.3\,(4) \\
 40 & 0.0061 & 68.7211\,(2)  & -40.802\,(9) & -27.739\,(4) & 0.19\,(1)   && -0.3\,(1) \\
 50 & -0.0294 & 37.6546\,(1) & -19.659\,(7) & -17.768\,(2) & 0.198\,(7)  && -0.24\,(5) \\
 60 & -0.0814 & 22.8815\,(1) & -10.405\,(5) & -12.192\,(2) & 0.203\,(6)  & 0.22\,(7) & -0.18\,(3) \\
 70 & -0.1524 & 15.0274      & -5.867\,(4)  & -8.814\,(3)  & 0.193\,(5)  & 0.19\,(6) & -0.17\,(2) \\
 83 & -0.2869 & 9.5618       & -2.992\,(3)  & -6.147\,(3)  & 0.135\,(4)  & 0.13\,(4) & -0.26\,(1) \\
 92 & -0.4343 & 7.3930       & -1.972\,(3)  & -4.968\,(2)  & 0.019\,(3)  & 0.01\,(3) & -0.444\,(8) \\
100 & -0.6512 & 6.0999       & -1.437\,(2)  & -4.224\,(3)  & -0.212\,(4) &-0.21\,(3) & -0.786\,(7) \\
\end{tabular}
\end{ruledtabular}
\end{table*}

The two-loop self-energy correction to energy levels is conveniently parameterized in terms of the
dimensionless function $F(\Za)$ defined as
\begin{align}
\label{FZa} \Delta E_{\rm SESE} &\ = m \left(\frac{\alpha}{\pi}\right)^2
\frac{(Z\alpha)^4}{n^3}\,F(\Za)\,.
\end{align}
In the present work we calculate the function $F(\Za)$ to all orders in $\Za$. In order to compare
our results with calculations based on the $\Za$ expansion, it is convenient to identify the
higher-order remainder function $G_{\rm h.o.}$ that incorporates contributions of all orders
starting with $\alpha^2(\Za)^6$,
\begin{align} \label{Gho}
F(\Za) =&\
B_{40}+ (Z\alpha)B_{50}
 \nonumber \\ &
+ (Z\alpha)^2 \Bigl[
  L^3 B_{63}
  +L^2 B_{62} +  L\,B_{61} + G_{\rm h.o.}(Z) \Bigr]
\,,
\end{align}
where  $L \equiv \ln[(Z\alpha)^{-2}]$  and the expansion of the remainder starts with a constant,
$G_{\rm h.o.}(Z) = B_{60}+ \Za \,(\ldots)\,$. Available results for the expansion coefficients
$B_{40}$-$B_{61}$ \cite{pachucki:01:pra,pachucki:03:prl,czarnecki:05:prl,jentschura:05:sese} are
summarized in Table~I of Ref.~\cite{yerokhin:15:Hlike}.

Our numerical results for the two-loop self-energy correction for the $1s$, $2s$, and $2p_{1/2}$
states of H-like ions with $Z = 30$-$100$ are presented in Table~\ref{tab:sese}. The last but one
column contains our previous results taken from Ref.~\cite{yerokhin:05:sese} for the $1s$ state and
from Ref.~\cite{yerokhin:06:prl} for excited states. As can be seen from the table, the numerical
accuracy was improved by an order of magnitude as compared to our previous works and calculations
for excited states were extended to the region $Z = 30$-$50$. Agreement with our previous
calculations is very good. A small deviation in the high-$Z$ region for the $1s$ state is due to
the fact that the $P$-term results in Ref.~\cite{yerokhin:05:sese} partly included the finite
nuclear size effect, whereas the present results are obtained strictly for the point nucleus.

It is remarkable that, while the total values of the function $F(\Za)$ are quite small numerically
(of order of $1$ for the $s$ states and of order of $0.1$ for the $p$ states), the individual
contributions listed in Tables~\ref{tab:mterm}-\ref{tab:pterm} are larger, often by orders of
magnitude. So, the final results for the two-loop self-energy are obtained through delicate
cancellations of numerous individual contributions to the $M$, $P$, and $F$ terms, the
cancellations growing fast as $Z$ decreases. Because of the strong $Z$-dependence of the
cancellations, an analysis of the final results for $F(\Za)$ and $G_{\rm h.o.}(Z)$ as functions of
$Z$ and a comparison with the $\Za$ expansion yields an independent check of correctness of the
results obtained and of our estimations of errors of the evaluation.

The analysis of the all-order two-loop self-energy results for the $1s$ state and the comparison
with the corresponding $\Za$-expansion coefficients were reported in our previous investigation
\cite{yerokhin:09:sese}. In this work we do not repeat this analysis since the improved numerical
accuracy for $Z \ge 30$ does not influence the extrapolation to $Z\to 0$. Instead, we present an
analysis for the normalized difference of the two-loop self-energy for the $1s$ and $2s$ states,
$\delta E \equiv 8\Delta E_{2s} - \Delta E_{1s}$. This difference is known within the $\Za$
expansion to a much better extent than $\Delta E_{2s}$ and $\Delta E_{1s}$ separately.
Specifically, there are results available for the first two expansion coefficients of the
higher-order remainder,
\begin{align}
\delta G_{\rm h.o.}(Z) = \delta B_{60} +  (\Za)\,\ln[(Z\alpha)^{-2}]\,\delta B_{71} + \ldots\,,
\end{align}
where $\delta B_{60} = 14.1\,(4)$  and $\delta B_{71} = 15.9 \pm 8.0$ \cite{jentschura:05:sese}.

\begin{figure*}
\centerline{
\resizebox{0.75\textwidth}{!}{%
  \includegraphics{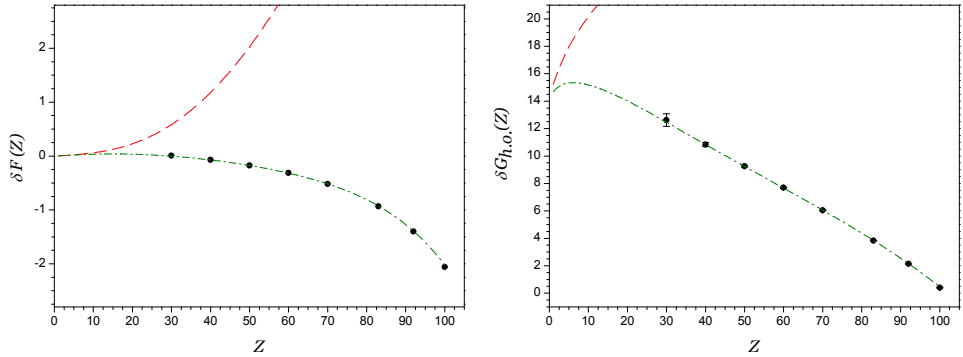}
}}
 \caption{The two-loop self-energy correction for the normalized difference of the $2s$ and $1s$ states, for the function
 $\delta F(\Za) = F_{2s}(\Za) - F_{1s}(\Za)$ (left graph) and for the higher-order remainder
 $\delta G_{\rm h.o.} = G_{\rm h.o.}(2s) - G_{\rm h.o.}(1s)$ (right graph). The dots denote the numerical all-order results, the
 dashed line (red) represents the $\Za$-expansion results, the dashed-dotted line (green) shows the best fit of the numerical data.
 \label{fig:sese2s1s}}
\end{figure*}

Figure~\ref{fig:sese2s1s} shows our all-order numerical results for the function $\delta F(\Za) =
F_{2s}(\Za) - F_{1s}(\Za)$ and the corresponding higher-order remainder $\delta G_{\rm h.o.}$, in
comparison with the $\Za$-expansion results. We observe that the $\Za$ expansion converges slowly
and that the known expansion coefficients are not sufficient in order to describe the all-order
results with $Z\ge 30$ even qualitatively. On the other hand, the numerical accuracy and the $Z$
range of the all-order results are not sufficient for extrapolating them to $Z\to 0$ and making a
clear statement about agreement with the $\Za$ expansion. Instead of this, we decided to assume the
correctness of the existing $\Za$-expansion results and to search for the best fitting function
that reconciles the $\Za$-expansion and all-order data. We found that we can describe our numerical
data very well by introducing just 3 fitting parameters. As a result, the best fit to our numerical
data is found to be
\begin{align}
\delta G_{\rm h.o., fit}(Z) = &\ \delta B_{60} +  (\Za)\,\biggl\{\ln[(Z\alpha)^{-2}]\,\delta B_{71}
 \nonumber \\ &
+ b_{70} + (\Za)\,b_{80} + (\Za)^2\,b_{90}\biggr\}\,,
\end{align}
where the fitted parameters are $b_{70} = -75.324$, $b_{80} = 100.336$, $b_{90} = -49.917$. The
fitted function is plotted in Fig.~\ref{fig:sese2s1s} with the dashed-dotted line. Our general
conclusion is that our all-order results for the normalized difference of the $2s$ and $1s$ states
are consistent with the available $\Za$-expansion coefficients.

\begin{figure*}
\centerline{
\resizebox{0.75\textwidth}{!}{%
  \includegraphics{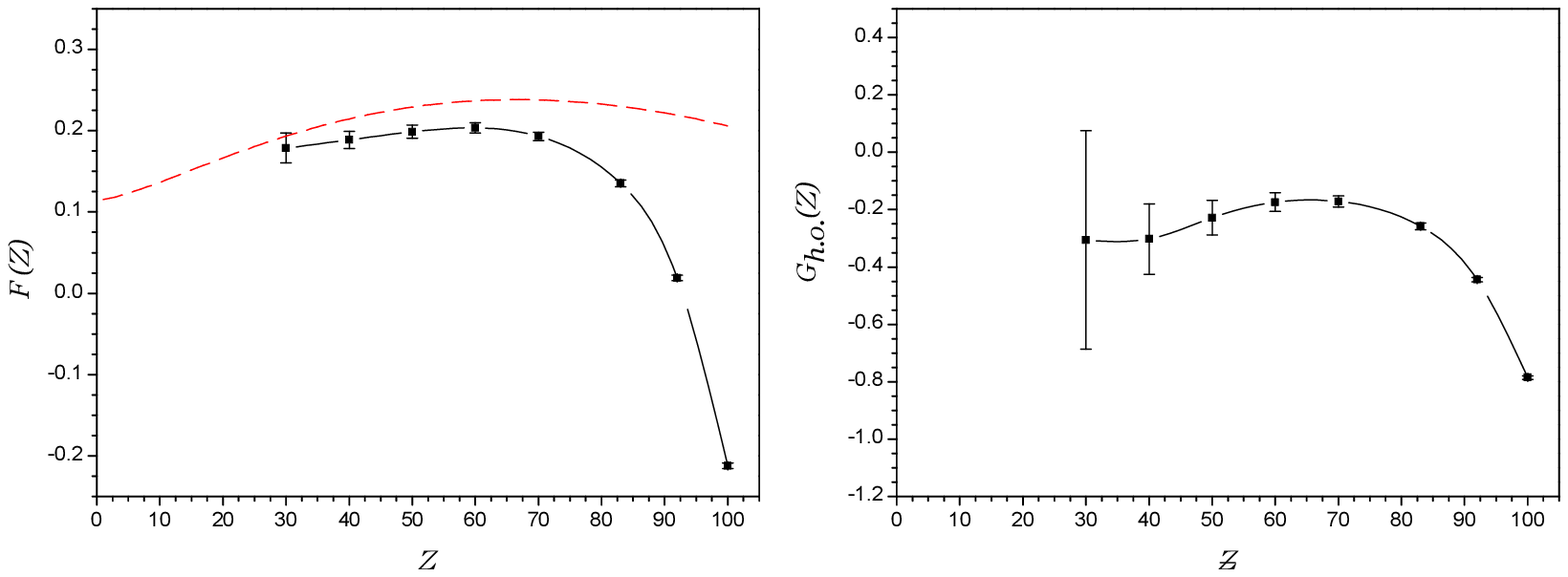}
}}
 \caption{The two-loop self-energy correction for the $2p_{1/2}$ state, for the function
 $F(\Za)$ (left graph) and for the higher-order remainder
 $G_{\rm h.o.}$ (right graph). The dots denote the numerical all-order results, whereas the
 dashed line (red) represents the $\Za$-expansion results.
 \label{fig:sese2p1}}
\end{figure*}

In Fig.~\ref{fig:sese2p1} we plot our numerical results for the $2p_{1/2}$ state, in comparison
with the corresponding $\Za$-expansion results. We observe that in this case the known terms of the
$\Za$ expansion qualitatively reproduce the behaviour of the all-order results for the function
$F(\Za)$ in the medium-$Z$ range. The numerical values of the higher-order remainder $G_{\rm h.o.}$
turn out to be rather small in this case. The accuracy of our all-order results is not sufficient
for the extrapolation of $G_{\rm h.o.}$ to $Z\to 0$, but we can conclude that there is a
qualitative agreement with the $\Za$ expansion for the $2p_{1/2}$ state.

Results for the two-loop self-energy correction for the $2p_{3/2}$ state were reported previously
for several high-$Z$ ions \cite{yerokhin:06:prl}. In the present work, we performed calculations
for the $2p_{3/2}$ state, the results being consistent with those from Ref.~\cite{yerokhin:06:prl}.
However, extending our calculations down till $Z = 30$, we found inconsistency with the
$\Za$-expansion results. We assume that this indicates an error in our codes for the $2p_{3/2}$
state, which we were not able to locate so far. For this reason we do not present any results for
the $2p_{3/2}$ state in this paper.

%
%
\section*{Conclusion}

We carried out calculations of the two-loop self-energy correction to the energy levels of the
$1s$, $2s$, and $2p_{1/2}$ states of hydrogen-like ions with the nuclear charges $Z = 30$-$100$.
The calculation was performed to all orders in the nuclear binding strength parameter $\Za$, for
the point distribution of the nuclear charge. The obtained results improved the accuracy of the
previously published values for the two-loop self-energy by an order of magnitude. For excited
states, they also extended the lowest nuclear charge range from $Z = 60$ to $Z = 30$. The obtained
results were shown to be consistent with the known coefficients of the $\Za$ expansion.

\section*{Acknowledgement}

This work was performed with support by the Ministry of Education and Science of the Russian
Federation Grant No.~3.5397.2017/6.7.


\begin{thebibliography}{10}

\bibitem{mohr:98} P.~J. Mohr, G.~Plunien, and G.~Soff,
\newblock Phys. Rep. {\bf 293}, 227  (1998).

\bibitem{shabaev:02:rep} V.~M. Shabaev,
\newblock Phys. Rep. {\bf 356}, 119  (2002).

\bibitem{yerokhin:15:Hlike} V.~A. Yerokhin and V.~M. Shabaev,
\newblock J. Phys. Chem. Ref. Data {\bf 44}, 033103 (2015).

\bibitem{fischer:04} M.~Fischer, N.~Kolachevsky, M.~Zimmermann, R.~Holzwarth, T.~Udem, T.~W.
  H\"ansch, M.~Abgrall, J.~Gr\"unert, I.~Maksimovic, S.~Bize, H.~Marion,
  F.~P.~D. Santos, P.~Lemonde, G.~Santarelli, P.~Laurent, A.~Clairon,
  C.~Salomon, M.~Haas, U.~D. Jentschura, and C.~H. Keitel,
\newblock Phys. Rev. Lett. {\bf 92}, 230802 (2004).

\bibitem{brandau:04} C.~Brandau, C.~Kozhuharov, A.~M\"uller, W.~Shi, S.~Schippers, T.~Bartsch,
  S.~B\"ohm, C.~B\"ohme, A.~Hoffknecht, H.~Knopp, N.~Gr\"un, W.~Scheid,
  T.~Steih, F.~Bosch, B.~Franzke, P.~H. Mokler, F.~Nolden, M.~Steck,
  T.~St\"ohlker, and Z.~Stachura,
\newblock Phys. Rev. Lett. {\bf 91}, 073202 (2003).

\bibitem{beiersdorfer:05} P.~Beiersdorfer, H.~Chen, D.~B. Thorn, and E.~Tr\"abert,
\newblock Phys. Rev. Lett. {\bf 95}, 233003 (2005).

\bibitem{pachucki:01:pra} K.~Pachucki,
\newblock Phys. Rev. A {\bf 63}, 042503 (2001).

\bibitem{pachucki:03:prl} K.~Pachucki and U.~D. Jentschura,
\newblock Phys. Rev. Lett. {\bf 91}, 113005 (2003).

\bibitem{czarnecki:05:prl} A.~Czarnecki, U.~D. Jentschura, and K.~Pachucki,
\newblock Phys. Rev. Lett. {\bf 95}, 180404 (2005).

\bibitem{jentschura:05:sese} U.~D. Jentschura, A.~Czarnecki, and K.~Pachucki,
\newblock Phys. Rev. A {\bf 72}, 062102 (2005).

\bibitem{yerokhin:01:prl} V.~A. Yerokhin,
\newblock Phys. Rev. Lett. {\bf 86}, 1990  (2001).

\bibitem{yerokhin:03:prl} V.~A. Yerokhin, P.~Indelicato, and V.~M. Shabaev,
\newblock Phys. Rev. Lett. {\bf 91}, 073001 (2003).

\bibitem{yerokhin:06:prl} V.~A. Yerokhin, P.~Indelicato, and V.~M. Shabaev,
\newblock Phys. Rev. Lett. {\bf 97}, 253004 (2006).

\bibitem{yerokhin:08:twoloop} V.~A. Yerokhin, P.~Indelicato, and V.~M. Shabaev,
\newblock Phys. Rev. A {\bf 77}, 062510 (2008).

\bibitem{yerokhin:09:sese} V.~A. Yerokhin,
\newblock Phys. Rev. A {\bf 80}, 040501(R) (2009).

\bibitem{yerokhin:05:sese} V.~A. Yerokhin, P.~Indelicato, and V.~M. Shabaev,
\newblock Phys. Rev. A {\bf 71}, 040101(R) (2005).

\bibitem{mohr:16:codata} P.~J. Mohr, D.~B. Newell, and B.~N. Taylor,
\newblock Rev. Mod. Phys. {\bf 88}, 035009 (2016).

\bibitem{pohl:05} R.~Pohl, A.~Antognini, F.~D. Amaro, F.~B. J. M.~R. Cardoso, C.~A.~N. Conde,
  A.~Dax, S.~Dhawan, L.~M.~P. Fernandes, T.~W. H\"ansch, F.~J. Hartmann, V.~W.
  Hughes, O.~Huot, P.~Indelicato, L.~Julien, P.~E. Knowles, F.~Kottmann, Y.-W.
  Liu, L.~Ludhova, C.~M.~B. Monteiro, F.~Mulhauser, F.~Nez, P.~Rabinowitz,
  J.~M.~F. dos Santos, L.~A. Schaller, C.~Schwob, D.~Taqqu, and J.~F. C.~A.
  Veloso,
\newblock Can. J. Phys. {\bf 83}, 339  (2005).

\bibitem{yerokhin:03:epjd} V.~A. Yerokhin, P.~Indelicato, and V.~M. Shabaev,
\newblock Eur. Phys. J. D {\bf 25}, 203  (2003).

\bibitem{mitrushenkov:95} A.~Mitrushenkov, L.~Labzowsky, I.~Lindgren, H.~Persson, and
    S.~Salomonson,
\newblock Phys. Lett. {\bf A200}, 51  (1995).

\bibitem{mallampalli:98:prl} S.~Mallampalli and J.~Sapirstein,
\newblock Phys. Rev. Lett. {\bf 80}, 5297  (1998).

\bibitem{yerokhin:00:lalpra} V.~A. Yerokhin,
\newblock Phys. Rev. A {\bf 62}, 012508 (2000).

\bibitem{indelicato:92:se} P.~Indelicato and P.~J. Mohr,
\newblock Phys. Rev. A {\bf 46}, 172 (1992).

\bibitem{indelicato:98} P.~Indelicato and P.~J. Mohr,
\newblock Phys. Rev. A {\bf 58}, 165 (1998).

\bibitem{indelicato:01} P.~Indelicato and P.~J. Mohr,
\newblock Phys. Rev. A {\bf 63}, 052507 (2001).

\bibitem{indelicato:14} P.~Indelicato, P.~J. Mohr, and J.~Sapirstein,
\newblock Phys. Rev. A {\bf 89}, 042121 (2014).

\bibitem{snyderman:91} N.~J. Snyderman,
\newblock Ann. Phys. (New York) {\bf 211}, 43 (1991).

\bibitem{mallampalli:98:pra} S.~Mallampalli and J.~Sapirstein,
\newblock Phys. Rev. A {\bf 57}, 1548  (1998).

\bibitem{yerokhin:01:sese} V.~A. Yerokhin and V.~M. Shabaev,
\newblock Phys. Rev. A {\bf 64}, 062507 (2001).

\bibitem{mohr:74:a} P.~J. Mohr,
\newblock Ann. Phys. (NY) {\bf 88}, 26  (1974).

\bibitem{yerokhin:05:se} V.~A. Yerokhin, K.~Pachucki, and V.~M. Shabaev,
\newblock Phys. Rev. A {\bf 72}, 042502 (2005).

\bibitem{pachucki:14:cpc} K.~Pachucki, M.~Puchalski, and V.~Yerokhin,
\newblock Comput. Phys. Commun. {\bf 185}, 2913  (2014).

\bibitem{johnson:88} W.~R. Johnson, S.~A. Blundell, and J.~Sapirstein,
\newblock Phys. Rev. A {\bf 37}, 307  (1988).

\bibitem{shabaev:04:DKB} V.~M. Shabaev, I.~I. Tupitsyn, V.~A. Yerokhin, G.~Plunien, and G.~Soff,
\newblock Phys. Rev. Lett. {\bf 93}, 130405 (2004).

\bibitem{yerokhin:10:sese} V.~A. Yerokhin,
\newblock Eur. Phys. J. D {\bf 58}, 57 (2010).

\bibitem{yerokhin:99:pra} V.~A. Yerokhin and V.~M. Shabaev,
\newblock Phys. Rev. A {\bf 60}, 800  (1999).

\bibitem{blundell:92} S.~A. Blundell,
\newblock Phys. Rev. A {\bf 46}, 3762  (1992).

\bibitem{blundell:91:se} S.~A. Blundell and N.~J. Snyderman,
\newblock Phys. Rev. A {\bf 44}, R1427  (1991).

\bibitem{johnson:88:b} W.~R.~Johnson, S.~A.~Blundell and J.~Sapirstein
\newblock Phys. Rev. A {\bf 37}, 2764 (1998).

\end{thebibliography}

\appendix

\section{Definitions, notations and useful identities}
\label{app:1}

The the photon propagator in the Feynman gauge is
\begin{align}
D_{\mu\nu}(\omega,x_{12}) = g_{\mu\nu}\,
D(\omega,x_{12}) \equiv g_{\mu\nu}\,\frac{\exp(i\, \sqrt{\omega^2+i0}\,x_{12})}{4\pi x_{12}}
\,,
\end{align}
where $x_{12} = |\bfx_1-\bfx_2|$, and the branch of the square root is fixed by the condition
${\rm{Im}}(\sqrt{\omega^2+ i 0})>0$.

The Green function of the Dirac-Coulomb equation is defined by its spectral representation
\begin{align}
G(\vare) = \sum_n \frac{|n\rbr\lbr n|}{\vare-\vare_{n}}\,,
\end{align}
where the summation over $n$ is performed over the complete spectrum of the Dirac equation with the
Coulomb nuclear potential. The free Dirac Green function is the $Z\to 0$ limit of the Dirac-Coulomb
Green function,
\begin{align}
G^{(0)}(\vare) = \sum_{\alpha } \frac{|\alpha \rbr\lbr \alpha |}{\vare-\vare_{\alpha }} = G(\vare) \bigr|_{Z = 0}\,.
\end{align}
Here and everywhere in this paper, greek subscripts $\alpha$, $\beta$, etc., refer to states of the
{\em free} electron, whereas italic subscripts $n$, $k$, etc., to states in the binding Coulomb
potential. The one-potential Dirac Green function is defined as the linear in $Z$ term of the $Z$
expansion of the Dirac-Coulomb Green function,
\begin{align}
G^{(1)}(\vare) = Z\, \left[ \frac{d}{dZ}\,G(\vare) \right]_{Z = 0}\,.
\end{align}
We introduce special notations for the Dirac Green function containing one and more (two and more)
interactions with the binding potential, defined as
\begin{align}
G^{(1+)}(\vare) &\ = G(\vare) - G^{(0)}(\vare)\,,\\
G^{(2+)}(\vare) &\  = G(\vare) - G^{(0)}(\vare)- G^{(1)}(\vare)\,.
\end{align}

We also mention the identity that follows from the Dirac equation, which is extensively used in
this paper. It reads, in coordinate space,
\begin{equation}\label{003a}
\psi_a(\bfx_1) = \int d\bfx_2 \,G^{(0)}(\vare_a,\bfx_1,\bfx_2)\,V(\bfx_2)\,\psi_a(\bfx_2)\,,
\end{equation}
where $V(\bfx)$ is the binding Coulomb potential, and in momentum space,
\begin{equation}\label{003b}
    \psi_a(\bfp) = \frac1{\gamma^0\vare_a - \bgamma\cdot\bfp-m} \,\gamma^0 \,(V
    \psi_a)(\bfp)\,,
\end{equation}
where $(V \psi_a)(\bfp)$ is the Fourier transform of the product $V(x)\,\psi_a(\bfx)$.

\section{Electron-electron interaction operator}
\label{app:2}

The operator of the electron-electron interaction in the Feynman gauge is
\begin{align}
I(\omega) &\ = e^2 \alpha^{\mu}\alpha_{\mu} D(\omega,x_{12})
 \nonumber \\
     &\ = \alpha \,(1-\balpha_1\cdot\balpha_2)\,\frac{\exp(\im
    \sqrt{\omega^2+\im 0}\,  x_{12})}{x_{12}}\,,
\end{align}
where $\alpha = e^2/(4\pi)$ is the fine structure constant and ${\alpha}^{\mu} = (1, \balpha) $ is
the vector of Dirac matrices.

The matrix elements of the electron-electron interaction operator is conveniently represented in
the form \cite{johnson:88:b,blundell:92}
\begin{equation} \label{Rdef}
\lbr ab| I(\omega)| cd \rbr
         =  \alpha\, \sum_{L} J_L(abcd)\, R_L(\omega,
         abcd)\,,
\end{equation}
where the function $J_L$ contains the standard magnetic-substate dependence of a scalar two-body
operator,
\begin{eqnarray}
J_L(abcd) &=& \sum_{m_L} \frac{(-1)^{L-m_L+j_c-\mu_c+j_d-\mu_d}}{2L+1}\,
  \nonumber \\ &&
  \times
C^{Lm_L}_{j_a\mu_a,j_c-\mu_c} \, C^{Lm_L}_{j_d\mu_d,j_b-\mu_b}\,,
\end{eqnarray}
$C^{jm}_{j_1\mu_1,\,j_2\mu_2}$ are the Clebsch-Gordan coefficients and $R_L$ is the relativistic
generalization of the Slater radial integral (for explicit expressions see, e.g.,
Ref.~\cite{blundell:92} and Appendix C of Ref.~\cite{yerokhin:03:epjd}). The sum over $L$ in
Eq.~(\ref{Rdef}) is restricted by the triangular selection rules of the Clebsch-Gordan
coefficients. It is noteworthy that the operator $I$ preserves the total momentum projection, i.e.,
the nonzero matrix elements should comply with the requirement
\begin{equation}
    \mu_a+\mu_b = \mu_c+\mu_d\,.
\end{equation}

\section{One-loop self-energy}
\label{app:se}

In this section we summarize the main definitions and notations for the one-loop self-energy
correction, which are extensively used throughout the paper.

The unrenormalized one-loop self-energy operator is given by
\begin{align} \label{sefunction}
\Sigma(\vare,\bfx_1,\bfx_2) = 2i\alpha\gamma^0 \int_{C_F} d\omega\,
    D(\omega,x_{12})\,
      \alpha_{\nu}
         G(\vare-\omega,\bfx_1,\bfx_2) \alpha^{\nu},
\end{align}
where $C_F$ is the standard Feynman integration contour. The renormalization of the one-loop
self-energy is performed \cite{snyderman:91,blundell:91:se,blundell:92} by expanding the
Dirac-Coulomb Green function $G$ in terms of the interaction with the binding Coulomb field. Using
the identity
\begin{align} \label{eqa01}
G(\vare) = G^{(0)}(\vare) + G^{(1)}(\vare) + G^{(2+)}(\vare)\,,
\end{align}
one represents the one-loop self-energy correction to the energy as a sum of the zero-potential,
one-potential, and many-potential terms, which are induced by the three terms in the
right-hand-side of Eq.~(\ref{eqa01}), correspondingly,
\begin{align}
\Delta E_{\rm SE} = \lbr a|\gamma_0\,\widetilde{\Sigma}(\vare_a)|a\rbr =   \Delta E^{\rm zero}_{\rm SE}+
        \Delta E^{\rm one}_{\rm SE}+ \Delta E^{\rm many}_{\rm SE} \,,
\end{align}
where $\widetilde{\Sigma}(\vare) = \Sigma(\vare)-\delta m$ and $\delta m$ is the corresponding mass
counterterm.

The zero-potential term is given by the matrix element of the renormalized free self-energy
operator $\Sigma_R^{(0)}(\vare)$ in momentum space,
\begin{align}\label{eq1a}
\Delta E^{\rm zero}_{\rm SE} &\ = \lbr a|\gamma^0\,\Sigma_R^{(0)}(\vare_a)|a\rbr
  \nonumber \\
&\ = \int
   \frac{d\bfp}{(2\pi)^3}\,\, \psi^{\dag}_{a}(\bfp)\,\gamma^0\,
     \Sigma_R^{(0)}(\vare_a,\bfp)\, \psi_{a}(\bfp)\,.
\end{align}
Explicit formulas for the operator $\Sigma_R^{(0)}(\vare)$ can be found, e.g., in Appendix~A of
Ref.~\cite{yerokhin:99:pra} for the case of $D=4$ dimensions and in Appendix~A.1 of
Ref.~\cite{yerokhin:03:epjd} for the general case of $D$ dimensions.

The one-potential term is given by the matrix element of the renormalized free vertex operator
$\Gamma_R^{0}(\vare_1,\vare_2)$ in momentum space,
\begin{align} \label{eq:onepot}
 &\ \Delta E^{\rm one}_{\rm SE} =
\lbr a|V\,\gamma^0\,\Gamma_R^{0}(\vare_a,\vare_a)|a\rbr
  \nonumber \\
&\ = \int \frac{d\bfp_1}{(2\pi)^3}\,\frac{d\bfp_2}{(2\pi)^3}\,\,
  \psi^{\dag}_{a}(\bfp_1)\,    V(\bfq) \, \gamma^0\,
    \Gamma_R^{0}(\vare_a,\bfp_1;\vare_a,\bfp_2)\,
    \psi_{a}(\bfp_2)\,,
\end{align}
where $V(\bfq) = -4\pi Z\alpha/|\bfq|$ is the Coulomb potential in the momentum space and  $\bfq =
\bfp_1-\bfp_2$. Explicit formulas for the renormalzed free vertex operator can be found, e.g., in
Appendix~B of Ref.~\cite{yerokhin:99:pra} for the case of $D=4$ dimensions and in Appendix~A.2 of
Ref.~\cite{yerokhin:03:epjd} for the general case of $D$ dimensions.

The many-potential term is given by the matrix element of the subtracted self-energy operator in
coordinate space,
\begin{align}
\Delta E^{\rm many}_{\rm SE} = &\ \lbr a|\gamma^0 \left[\Sigma(\vare_a) - \Sigma^{(0)}(\vare_a)
- \Sigma^{(1)}(\vare_a)\right] |a\rbr
  \nonumber \\
 = &\ 2\,i\,\alpha\,\int_{C_F} d\omega
        \int d\bfx_1\,
        d \bfx_2\, D(\omega, \bfx_{12})\,
        \psi^{\dag}_a(\bfx_1)\,
  \nonumber \\  & \times
        \alpha_{\nu}\,
        G^{(2+)}(\vare_a-\omega,\bfx_1,\bfx_2)\, \alpha^{\nu}\,
        \psi_a(\bfx_2)\, .
\end{align}

In order to bring the many-potential term to the form suitable for a numerical evaluation, one
needs to perform the integrations over the angular variables, sum over the angular-momentum
projections, and deform the integration contour. The result \cite{yerokhin:99:pra} can be written
down as
\begin{align}
\Delta E^{\rm many}_{\rm SE} = \frac{i\alpha}{2\pi}
   \int_{C_{LH}} d\omega\,
   &\
   \Biggl[ \sum_{nJ} \frac{(-1)^{J+j_n-j_a}}{2j_a+1}\,
       \nonumber \\ & \times
    \frac{R_J(\omega,anna)}{\vare_a-\omega-\vare_n}
    - \mbox{\rm Subtractions} \Biggr] \,,
\end{align}
where the subtractions are defined symbolically by the substitution
\begin{align}
G(\vare) \to G^{(2+)}(\vare)\,,
\end{align}
and the integration contour $C_{LH}$ is shown on Fig.~\ref{fig:CLH}. Specifically, the contour
$C_{LH}$ consists of the low-energy part $C_L$ and the high-energy part $C_H$. The high-energy part
$C_H$ is parallel to the imaginary axis and extends from $\Delta-i\infty$ to $\Delta$ and from
$\Delta$ to $\Delta+i\infty$. Such a choice of the contour eliminates strong oscillations of the
integrand arising in the high-energy region of the contour $C_F$ and replaces them by the
exponential falling-off. The low-energy part of the integration contour $C_L$ runs over the upper
and the lower banks of the cut of the photon propagator. In the general case of excited reference
states, it is also bent in the complex plane in order to avoid singularities coming from virtual
bound states with energies $\vare_n < \vare_a$ in the electron propagator. Specifically, the
contour $C_L$ consists of the upper and lower parts, both of which extend over 3 sections:
$[0,\delta_{x,1}-i\delta_y]$, $[\delta_{x,1}-i\delta_y,\delta_{x,2}]$, and $[\delta_{x,2},\Delta]$,
as shown on Fig.~\ref{fig:CLH}. The advantage of such a choice of the low-energy part of the
contour is that virtual bound states with the energy $\vare_n \leq \vare_a$ in the electron
propagator do not create any pole contributions and do not require any special treatment. The
parameters of the contour $\delta_{x,1}$, $\delta_{x,2}$, $\delta_{y}$, and $\Delta$ may be chosen
differently. In our recent works, we used the following choice: $\delta_{x,1} = \vare_a -
\vare_{1s}$ when the reference state $a$ is an excited state and $\delta_{x,1} = (\Za)^2$ when $a$
is the $1s$ state; $\delta_{x,2} = 2\,\delta_{x,1}$; $\delta_y = \delta_{x,1}/2$ when there are
intermediate states with the energy $\vare_n < \vare_a$ and $\delta_y = 0$ otherwise; and $\Delta =
\Za\,\vare_a$.

We note that for the case when the reference state is the ground state, there is no need to bend
the low-energy part of the contour in the complex plane (as there are no intermediate states with
energy $0 <\vare_n < \vare_a$), so it is convenient to set $\delta_y = 0$.

\end{document}